\documentclass{article}
\usepackage{graphicx}
\usepackage[centertags]{amsmath}
\usepackage{amsfonts}
\usepackage{amssymb}
\usepackage{amsthm}

\title{Quasiclassical and Quantum Systems of Angular Momentum. Part III. Group Algebra of ${\rm SU}(2)$, Quantum Angular Momentum and Quasiclassical Asymptotics}

\author{J. J. S\l awianowski, V. Kovalchuk, A. Martens,\\ 
B. Go\l ubowska, and E. E. Ro\.zko\\
Institute of Fundamental Technological Research,\\
Polish Academy of Sciences,\\
$5^{\rm B}$, Pawi\'{n}skiego str., 02-106 Warsaw, Poland\\
e-mails: jslawian@ippt.gov.pl, vkoval@ippt.gov.pl,\\ 
amartens@ippt.gov.pl, bgolub@ippt.gov.pl, erozko@ippt.gov.pl}

\begin{document}

\maketitle
\begin{abstract}
This is the third part of our series "Quasiclassical and Quantum Systems of Angular Momentum". In two previous parts we have discussed the methods of group algebras in formulation of quantum mechanics and certain quasiclassical problems. Below we specify to the special case of the group ${\rm SU}(2)$ and its quotient ${\rm SO}(3,\mathbb{R})$, and discuss just our main subject in this series, i.e., angular momentum problems. To be more precise, this is the purely ${\rm SU}(2)$-treatment, so formally this might also apply to isospin. However. it is rather hard to imagine realistic quasiclassical isospin problems.
\end{abstract}
  
Theory of angular momentum is based on the group ${\rm SU}(2)$ and its quotient ${\rm SO}(3,\mathbb{R})={\rm SU}(2)/\mathbb{Z}_{2}$. The two-element center and maximal normal divisor $\mathbb{Z}_{2}$ of the simply-connected group ${\rm SU}(2)$ is given by
\begin{equation}
\mathbb{Z}_{2}= \left\{I_{2},-I_{2}\right\}, 
\label{eq_3.1}
\end{equation}
where, obviously, $I_{2}$ is the $2 \times 2$ unit matrix. 

Let $\sigma_{a}$, $a=1,2,3$ denote Pauli matrices in the following convention:
\begin{equation}
\sigma_{1}= \left[
\begin{array}{cc}
	0 & 1\\
	1 & 0
\end{array}\right], \qquad \sigma_{2}= \left[
\begin{array}{cc}
	0 & -i\\
	i & 0
\end{array}\right], \qquad \sigma_{3}= \left[
\begin{array}{cc}
	1 & 0\\
	0 & -1
\end{array}\right]. 
\label{eq_3.2}
\end{equation}
They are basic traceless Hermitian $2 \times 2$-matrices. The Lie algebra of ${\rm SU}(2)$, ${\rm SU}(2)'$, consists of anti-Hermitian traceless matrices; the basic ones are chosen as
\begin{equation}
e_{a}:=\frac{1}{2i}\sigma_{a}. 
\label{eq_3.3}
\end{equation}
The corresponding structure constants are given by the Ricci symbol, more precisely,
\begin{equation}
\left[e_{a},e_{b}\right]= \varepsilon^{c}{}_{ab}e_{c}, 
\label{eq_3.4}
\end{equation}
where $\varepsilon_{abc}$ is just the totally antisymmetric Ricci symbol, $\varepsilon_{123}=1$, and the raising/lowering of indices is meant here in the sense of the "Kronecker delta" $\delta_{ab}$ as the standard metric of $\mathbb{R}^{3}$. So, this shift of indices is here analytically a purely "cosmetic" procedure, however we use it to follow the standard convention.

We know that ${\rm SU}(2)$ is the universal $2:1$ covering group of ${\rm SO}(3,\mathbb{R})$, the proper orthogonal group in $\mathbb{R}^{3}$. The projection epimorphism 
\begin{equation}\label{eq_3.4a}
{\rm SU}(2)\ni u \mapsto R = pr(u) \in SO(3,\mathbb{R})
\end{equation}
is given by
\begin{equation}
ue_{b}u^{-1}=ue_{b}u^{+}= e_{a}R^{a}{}_{b}. 
\label{eq_3.5}
\end{equation}

With respect to the basis (\ref{eq_3.3}) the Killing metric $\gamma$ has the components
\begin{equation}
\gamma_{ab}= - 2\delta_{ab}; 
\label{eq_3.6}
\end{equation}
obviously, the negative definiteness is due to the compactness of the simple algebra/group ${\rm SU}(2)'/{\rm SU}(2)$. For practical purposes one eliminates the factor $(-2)$ and takes the metric
\begin{equation}
\Gamma_{ab}= - \frac{1}{2}\gamma_{ab}= \delta_{ab}. 
\label{eq_3.7}
\end{equation}

In terms of the canonical coordinates of the first kind:
\begin{equation}
u(\overline{k})= \exp \left(k^{a}e_{a}\right)= \cos \frac{k}{2}I_{2}- \frac{i}{k} \sin \frac{k}{2}k^{a}\sigma_{a}, \label{eq_3.8}
\end{equation}
where, obviously, $k$ denotes the Euclidean length of the vector $\overline{k} \in \mathbb{R}^{3}$:
\begin{equation}
k= \sqrt{\overline{k} \cdot \overline{k}}= \sqrt{\delta_{ab} k^{a} k^{b}}. \label{eq_3.9}
\end{equation}
Its range is $[0,2\pi]$ and the range of the unit vector (versor) $\overline{n}:= \frac{\overline{k}}{k}$ is the total unit sphere $S^{2}(0,1)\subset \mathbb{R}^{3}$. This coordinate system is singular at $k=0$, $k=2\pi$, where
\begin{equation}
u(0\overline{n})= I_{2}, \qquad u(2\pi \overline{n})= -I_{2} 
\label{eq_3.10}
\end{equation}
for any $\overline{n} \in S^{2}(0,1)$. Obviously, the formula (\ref{eq_3.8}) remains meaningful for $k > 2\pi$, however, the "former" elements of ${\rm SU}(2)$ are then repeated.

Sometimes one denotes
\begin{equation}
\sigma_{0}=I_{2}, \qquad e_{0}= \frac{1}{2} I_{2}. 
\label{eq_3.11}
\end{equation}
Then (\ref{eq_3.8}) may be written down as follows:
\begin{equation}
u= \xi^{\mu}(\overline{k})\left(2e_{\mu}\right), 
\label{eq_3.12}
\end{equation}
where the summation convention is meant over $\mu = 0,1,2,3$,
\begin{equation}
\left(\xi^{0}\right)^{2}+ \left(\xi^{1}\right)^{2}+ \left(\xi^{2}\right)^{2}+ \left(\xi^{3}\right)^{2}=1, \label{eq_3.13}
\end{equation}
and this formula together with the structure of parametrization (\ref{eq_3.8}), (\ref{eq_3.12}) tells us that ${\rm SU}(2)$ is the unit sphere $S^{3}(0,1)$ in $\mathbb{R}^{4}$. Roughly speaking, $k=0$ is the "north pole" and $k=2 \pi$ is the corresponding "south pole".

This "pseudo-relativistic" notation is rather misleading. The point is that the matrices $\sigma_{\mu}$, $e_{\mu}$ above are used to represent linear mappings in $\mathbb{C}^{2}$, i.e., mixed tensors in $\mathbb{C}^{2}$. In the relativistic theory of spinors, e.g., in Lagrangians for (anti)neutrino fields, $\sigma_{\mu}$ are used as matrices of sesqulinear Hermitian forms, thus, twice covariant tensors on $\mathbb{C}^{2}$. The space of such forms carries an intrinsic conformal-Minkowskian structure (Minkowskian up to the normalization of the scalar product). And then $\sigma_{\mu}$ form a Lorentz-ruled multiplet. This is seen in the standard procedure of using ${\rm SL}(2,\mathbb{C})$ as the universal covering of the restricted Lorentz group ${\rm SO}(1,3)^{\uparrow}$, namely,
\begin{equation}
a \sigma_{\mu} a^{+}= \sigma_{\nu} \Lambda^{\nu}{}_{\mu} 
\label{eq_3.14}
\end{equation}
describes the covering assignment 
\begin{equation}\label{eq_3.14a}
{\rm SL}(2,\mathbb{C}) \ni a \mapsto \Lambda \in{\rm SO}(1,3)^{\uparrow}. 
\end{equation}
The four-dimensional quantity $\left(\xi^{0}, \xi^{1}, \xi^{2}, \xi^{3}\right)$ in (\ref{eq_3.12}), (\ref{eq_3.13}) may be also interpreted in terms of the group ${\rm SO}(4,\mathbb{R})$ and its covering group, however, this interpretation is relatively complicated and must not be confused with the relativistic aspect of the quadruplet of $\sigma_{\mu}$-matrices as analytical representants of sesquilinear forms.

The Lie algebra of ${\rm SO}(3,\mathbb{R})$, ${\rm SO}(3,\mathbb{R})'$, consists of $3 \times 3$ skew-symmetric matrices with real entries. The standard choice of basis of ${\rm SO}(3,\mathbb{R})'$, adapted to (\ref{eq_3.3}) and to the procedure (\ref{eq_3.5}), is given by matrices $E_{a}$, $a=1,2,3$, with entries
\begin{equation}
\left(E_{a}\right)^{b}{}_{c}:= - \varepsilon_{a}{}^{b}{}_{c}, 
\label{eq_3.15}
\end{equation}
where again $\varepsilon_{abc}$ is the totally antisymmetric Ricci symbol, and indices are "cosmetically" shifted with the help of the Kronecker symbol. Then, of course,
\begin{equation}
\left[E_{a}, E_{b}\right] = \varepsilon^{c}{}_{ab}E_{c}. 
\label{eq_3.16}
\end{equation}

In spite of having isomorphic Lie algebras, the groups ${\rm SU}(2)$, ${\rm SO}(3,\mathbb{R})\simeq{\rm SU}(2)/\mathbb{Z}_{2}$ are globally different. The main topological distinction is that ${\rm SU}(2)$ is simply connected and ${\rm SO}(3,\mathbb{R})$ is doubly connected.

Using canonical coordinates of the first kind, we have in analogy
to (\ref{eq_3.8}) the formula
\begin{equation}
R(\overline{k})= \exp \left(k^{a} E_{a}\right). 
\label{eq_3.17}
\end{equation}

Because of the obvious reasons, known from elementary geometry and mechanics, $\overline{k}$ is referred to as the rotation vector, $k= \sqrt{\overline{k} \cdot \overline{k}}$ is the rotation angle, and the unit vector (versor)
\begin{equation}
\overline{n}= \frac{\overline{k}}{k} 
\label{eq_3.18}
\end{equation}
is the oriented rotation axis. We use the all standard concepts and symbols of the vector calculus in $\mathbb{R}^{3}$, in particular, scalar products $\overline{a} \cdot \overline{b}$ and vector products $\overline{a} \times \overline{b}$. The rotation angle $k$ runs over the range $[0, \pi]$ and the antipodal points on the sphere $S^{2}(0, \pi) \subset \mathbb{R}^{3}$ are identified, they describe the same rotation,
\begin{equation}
R\left(\pi \overline{n}\right)= R\left(- \pi \overline{n}\right). 
\label{eq_3.19}
\end{equation}
Therefore, this sphere, taken modulo the antipodal identification, is the manifold of non-trivial square roots of the identity $I_{3}$ in ${\rm SO}(3,\mathbb{R})$. It is seen in this picture that ${\rm SO}(3,\mathbb{R})$ is doubly connected, because any curve in the ball $K^{2}(0,1) \subset \mathbb{R}^{3}$ joining two antipodal points on the boundary $S^{2}(0,1)$ is closed under this identification, i.e., it is a loop, but it cannot be continuously contracted into a single point. 

Obviously, formally, the values $k > \pi$ are admitted, however, they correspond to rotations by $k < \pi$, taken earlier into account. By abuse of language, in ${\rm SU}(2)$ the quantities $\overline{k}$, $k$ are also referred to as the rotation vector and rotation angle. But one must "rotate" by $4 \pi$ to go back to the same situation, not by $2 \pi$. The matrix of $R(\overline{k})$ is given by
\begin{equation}
R\left(\overline{k}\right)^{a}{}_{b}= \cos k \delta^{a}{}_{b}+ \frac{1}{k^{2}}\left(1-\cos k \right)k^{a}k_{b}+ \frac{1}{k} \sin k \varepsilon^{a}{}_{bc} k^{c}, \label{eq_3.20}
\end{equation}
i.e.,
\begin{equation}
R\left(\overline{k}\right)\overline{x}= \cos k \overline{x}+ \frac{1-\cos k}{k^{2}}\left(\overline{k} \cdot \overline{x} \right)\overline{k}+ \frac{\sin k}{k} \overline{k} \times \overline{x} , \label{eq_3.21}
\end{equation}
or, symbolically,
\begin{equation}
R(\overline{k})\cdot\overline{x}= \overline{x}+ \overline{k} \times \overline{x}+ \frac{1}{2!}\overline{k} \times \left(\overline{k} \times \overline{x} \right)+ \cdots + \frac{1}{n!} \overline{k} \times \left(\overline{k} \times \cdots \times \left(\overline{k} \times \overline{x}\right) \cdots \right)+ \ldots \label{eq_3.22}
\end{equation}

Let us distinguish between two ways of viewing, representing geometry of ${\rm SU}(2)$ and ${\rm SO}(3,\mathbb{R})$ in terms of some subsets in $\mathbb{R}^{3}$ as the space of rotation vectors $\overline{k}$ or, alternatively, in terms of closed submanifolds and their quotients in $\mathbb{R}^{4}$. 

As seen from (\ref{eq_3.13}), ${\rm SU}(2)$ is a unit sphere $S^{3}(0,1) \subset \mathbb{R}^{4}$, ${\rm SO}(3,\mathbb{R})$ is obtained by the antipodal identification. Then ${\rm SO}(3,\mathbb{R})$ is doubly connected because the curves on $S^{3}(0,1)$ joining antipodal points project to the quotient manifold onto closed loops non-contractible to points in a continuous way. In $\mathbb{R}^{3}$ the group ${\rm SU}(2)$ is represented by the ball $K^{2}(0,2 \pi)$; the whole shell $S^{2}(0,2 \pi)$ represents the single point $-I_{2} \in{\rm SU}(2)$. Then ${\rm SO}(3,\mathbb{R})$ is pictured as the ball $K^{2}(0,\pi) \subset \mathbb{R}^{3}$ with the antipodal identification of points on the shell $S^{2}(0, \pi)$, cf. (\ref{eq_3.19}). This exhibits the identification of ${\rm SO}(3,\mathbb{R})$ with the projective space $P\mathbb{R}^{3}$; the antipodally identified points on $S^{2}(0, \pi)$ represent the improper points at infinity in $\mathbb{R}^{3}$.

For certain reasons, both practical and deeply geometrical, it is convenient to use also another parametrization of ${\rm SO}(3,\mathbb{R})$, using so-called vector of final rotation:
\begin{equation}
\overline{\varkappa}= \frac{2}{k}\ {\rm tg}\ \frac{k}{2}\ \overline{k}; 
\label{eq_3.17a}
\end{equation}
obviously, in the neighbourhood of group identity, when $\overline{k} \approx 0$, $\overline{\varkappa}$ differs from $\overline{k}$ by higher-order quantity. The practical advantage of $\overline{\varkappa}$ is that the composition rule and the action of rotations are described by very simple and purely algebraic expressions:
\begin{equation}
R\left(\overline{\varkappa}_{1}\right) R\left(\overline{\varkappa}_{2}\right)= R\left(\overline{\varkappa}\right), \label{eq_3.17b}
\end{equation}
where
\begin{eqnarray}
\overline{\varkappa}&=& \left(1- \frac{1}{4}\overline{\varkappa}_{1} \cdot \overline{\varkappa}_{2}\right)^{-1} \left(\overline{\varkappa}_{1}+ \overline{\varkappa}_{2}+ \frac{1}{2}\overline{\varkappa}_{1} \times \overline{\varkappa}_{2}\right), \label{eq_3.17c}\\
R\left[\overline{\varkappa}\right] \overline{x}&=& \overline{x}+ \left(1+ \frac{1}{4}\overline{\varkappa}^{2} \right)^{-1} \overline{\varkappa} \times \left(\overline{x}+ \frac{1}{2}\overline{\varkappa} \times \overline{x}\right). \label{eq_3.17d}
\end{eqnarray}
An important property of this parametrization is that it describes the projective mapping of ${\rm SO}(3,\mathbb{R})$ onto the projective space 
$P\mathbb{R}^{3}$. The one-parameter subgroups and their cosets in ${\rm SO}(3,\mathbb{R})$ are mapped onto straight-lines in $\mathbb{R}^{3}$. The manifold of $\pi$-rotations (non-trivial square roots of identity) is mapped onto the set of improper points in $P\mathbb{R}^{3}$, i.e., it "blows up" to infinity.

The homomorphism (\ref{eq_3.5}) of ${\rm SU}(2)$ onto ${\rm SO}(3,\mathbb{R})$, $u \mapsto R(u)$, may be alternatively described in terms of inner automorphisms of ${\rm SU}(2)$ and the rotation-vector parametrization:
\begin{equation}
u v(\overline{k})u^{-1}= v\left(R(u)\overline{k}\right),\qquad u \in{\rm SU}(2). 
\label{eq_3.23}
\end{equation}
Roughly speaking, inner automorphisms in ${\rm SU}(2)$ result in rotation of the rotation vector. Obviously, the same holds in ${\rm SO}(3,\mathbb{R})$:
\begin{equation}
O R(\overline{k})O^{-1}= R\left(O\overline{k}\right), \qquad O \in {\rm SO}(3,\mathbb{R}). \label{eq_3.24}
\end{equation}

Therefore, inner automorphisms preserve the length $k$ of the rotation vector $\overline{k}$, and the classes of conjugate elements are characterized by the fixed values of the rotation angle (but all possible oriented rotation axes $\overline{n}$). This means that in the above description they are represented by spheres $S^{2}(0,k) \subset \mathbb{R}^{3}$ in the space of rotation vectors. There are two one-element singular equivalence classes in ${\rm SU}(2)$, namely $\left\{I_{2}\right\}$, $\left\{-I_{2}\right\}$ corresponding respectively to $k=0$, $k=2 \pi$. Obviously, in ${\rm SO}(3,\mathbb{R})$ there is only one singular class $\left\{I_{3}\right\}$. More precisely, in ${\rm SO}(3,\mathbb{R})$ the class $k= \pi$ is not the sphere, but rather its antipodal quotient, so-called elliptic space. The idempotents $\varepsilon(\alpha)$/characters $\chi(\alpha)= \varepsilon(\alpha)/n(\alpha)$ and all central functions of the group algebras of ${\rm SU}(2)$ and ${\rm SO}(3,\mathbb{R})$ are constant on the spheres $S^{2}(0,k)$, i.e., depend on $\overline{k}$ only through the rotation angle $k$. In many problems it is convenient to parametrize ${\rm SU}(2)$ and ${\rm SO}(3,\mathbb{R})$ with the help of spherical variables $k$, $\theta$, $\varphi$ in the space $\mathbb{R}^{3}$ of the rotation vector $\overline{k}$. Historically the most popular parametrization is that based on the Euler angles $(\varphi,\vartheta, \psi)$. It is given by
\begin{eqnarray}
u[\varphi, \vartheta, \psi]= u(0, 0,\varphi)u(0, \vartheta, 0)u(0, 0, \psi),\label{eq_3.25a}  \\ 
R[\varphi, \vartheta, \psi]= R(0, 0,\varphi)R(0, \vartheta, 0)R(0, 0, \psi); \label{eq_3.25b}
\end{eqnarray}
historically $\varphi$, $\vartheta$, $\psi$ are referred to respectively as the precession angle, nutation angle and the rotation angle.

Sometimes one uses $u(\vartheta, 0, 0)$, $R(\vartheta, 0, 0)$ instead $u(0, \vartheta,  0)$, $R(0, \vartheta, 0)$ in (\ref{eq_3.25a}), (\ref{eq_3.25b}). The only thing which matters here is that one uses the product of three elements which belong to two one-parameter subgroups. The Euler angles are practically important in gyroscopic problems. Canonical parametrization of the second kind, 
\begin{equation}
u(\alpha, \beta, \gamma)= u(\alpha, 0, 0)u(0, \beta, 0)u(0, 0, \gamma) \label{eq_3.26}
\end{equation}
are not very popular; one must say, however, that many formulas have the same form in variables $(\varphi, \vartheta, \psi)$ and $(\alpha, \beta, \gamma)$.

It is well known that in ${\rm SU}(2)$ irreducible unitary representations, or rather their equivalence classes, are labelled by non-negative integers and half-integers,
\begin{equation}
\alpha = j = 0, \frac{1}{2}, 1, \frac{3}{2} ,\ldots,
\label{eq_3.27}
\end{equation}
i.e., $\Omega = \{0\}\bigcup \mathbb{N}/2$, where $\mathbb{N}$ denotes the set of naturals (positive integers). And, obviously,
\begin{equation}
n(\alpha)=n(j)=2j+1. \label{eq_3.28}
\end{equation}

On ${\rm SO}(3,\mathbb{R})$ one uses integers only,
\begin{equation} 
\alpha = j = 0, 1, 2, \ldots,\qquad \Omega = \{0\}\bigcup \mathbb{N}. \label{eq_3.29}
\end{equation}
For any $\alpha = j$, there is only one irreducible representation of dimension $n(\alpha)=n(j)=2j+1$; obviously, only one up to equivalence. It is not the case for many practically important groups, e.g., for ${\rm SU}(3)$ or for the non-compact group ${\rm SL}(2,\mathbb{C})$.

Historically, irreducible representations of ${\rm SU}(2)$, ${\rm SO}(3,\mathbb{R})$, ${\rm SL}(2,\mathbb{C})$, and ${\rm SO}(1,3)^{\uparrow}$ were found in two alternative ways:
\begin{itemize}
	\item[$(i)$] algebraic one, based on taking the tensor products of fundamental representation (by itself), 
	\item[$(ii)$] differential one, based on solving differential equations like (97)--(102), (109), (110) in \cite{2}.
\end{itemize}

The left and right generators ${\mathcal{L}}_{a}$, ${\mathcal{R}}_{a}$, i.e., respectively the basic right- and left-invariant vector fields, are analytically given by
\begin{eqnarray}
{\mathcal{L}}_{a}&=& \frac{k}{2} {\rm ctg} \frac{k}{2}\frac{\partial}{\partial k^{a}}+ \left(1- \frac{k}{2} {\rm ctg} \frac{k}{2} \right) \frac{k_{a}}{k} \frac{k^{b}}{k}\frac{\partial}{\partial k^{b}} + \frac{1}{2} \varepsilon_{ab}{}^{c}k^{b}\frac{\partial}{\partial k^{c}},\label{eq_3.30a}\\
{\mathcal{R}}_{a}&=& \frac{k}{2} {\rm ctg} \frac{k}{2}\frac{\partial}{\partial k^{a}}+ \left(1- \frac{k}{2} {\rm ctg} \frac{k}{2} \right) \frac{k_{a}}{k} \frac{k^{b}}{k}\frac{\partial}{\partial k^{b}} - \frac{1}{2} \varepsilon_{ab}{}^{c}k^{b}\frac{\partial}{\partial k^{c}},\label{eq_3.30b}
\end{eqnarray}
and therefore 
\begin{equation}
{\mathcal{A}}_{a}= {\mathcal{L}}_{a} - {\mathcal{R}}_{a} = \varepsilon_{ab}{}^{c}k^{b}\frac{\partial}{\partial k^{c}}. \label{eq_3.31}
\end{equation}

In terms of explicitly written components
\begin{eqnarray}
{\mathcal{L}}^{i}{}_{a}&=& \frac{k}{2} {\rm ctg} \frac{k}{2}\delta^{i}{}_{a} + \left(1- \frac{k}{2} {\rm ctg} \frac{k}{2} \right) \frac{k_{a}}{k} \frac{k^{i}}{k} + \frac{1}{2} \varepsilon_{ab}{}^{i}k^{b},\label{eq_3.32a} \\
{\mathcal{R}}^{i}{}_{a}&=& \frac{k}{2} {\rm ctg} \frac{k}{2} \delta^{i}{}_{a} + \left(1- \frac{k}{2} {\rm ctg} \frac{k}{2} \right) \frac{k_{a}}{k} \frac{k^{i}}{k} - \frac{1}{2} \varepsilon_{ab}{}^{i}k^{b},\label{eq_3.32b} \\
{\mathcal{A}}^{i}{}_{a}&=& \varepsilon_{ab}{}^{i}k^{b};\label{eq_3.32c}
\end{eqnarray}
obviously we mean here the shift of indices in the Kronecker-delta sense.

The corresponding Cartan one-forms are given by
\begin{eqnarray}
{\mathcal{L}}^{a}&=& \frac{\sin k}{k} dk^{a} + \left(1- \frac{\sin k}{k} \right) \frac{k^{a}}{k} \frac{k_{b}}{k}dk^{b} + \frac{2}{k^{2}} \sin^{2}\frac{k}{2} \varepsilon^{a}{}_{bc}k^{b}dk^{c},\label{eq_3.33a} \\
{\mathcal{R}}^{a}&=& \frac{\sin k}{k} dk^{a} + \left(1- \frac{\sin k}{k} \right) \frac{k^{a}}{k} \frac{k_{b}}{k}dk^{b} - \frac{2}{k^{2}} \sin^{2}\frac{k}{2} \varepsilon^{a}{}_{bc}k^{b}dk^{c},\label{eq_3.33b}
\end{eqnarray}
i.e., in terms of the components,
\begin{eqnarray}
{\mathcal{L}}^{a}{}_{i}&=& \frac{\sin k}{k} \delta^{a}{}_{i} + \left(1- \frac{\sin k}{k} \right) \frac{k^{a}}{k} \frac{k_{i}}{k} + \frac{2}{k^{2}} \sin^{2}\frac{k}{2} \varepsilon^{a}{}_{bi}k^{b},\label{eq_3.34a} \\
{\mathcal{R}}^{a}{}_{i}&=& \frac{\sin k}{k} \delta^{a}{}_{i} + \left(1- \frac{\sin k}{k} \right) \frac{k^{a}}{k} \frac{k_{i}}{k} - \frac{2}{k^{2}} \sin^{2}\frac{k}{2} \varepsilon^{a}{}_{bi}k^{b}.\label{eq_3.34b}
\end{eqnarray}

The central functions on ${\rm SU}(2)$ and on ${\rm SO}(3,\mathbb{R})$, in particular the idempotents $\varepsilon(j)$/characters $\chi(j)$ satisfy the obvious differential equations:
\begin{equation}
{\mathcal{A}}_{a}f=0, \qquad \text{i.e.,} \qquad {\mathcal{L}}_{a}f={\mathcal{R}}_{a}f, \qquad a=1,2,3. \label{eq_3.35}
\end{equation}

Obviously, the analytical formulas (\ref{eq_3.30a})-(\ref{eq_3.34b}) are formally valid both on ${\rm SU}(2)$ and ${\rm SO}(3, \mathbb{R})$, and in general the calculus on ${\rm SU}(2)$ is simpler than that on ${\rm SO}(3, \mathbb{R})$. It is convenient to rewrite the formulas (\ref{eq_3.30a})-(\ref{eq_3.34b}) so as to express them explicitly in terms of the angular and radial differential operations in the space of rotation vectors $\overline{k}$. After simple calculations one obtains
\begin{eqnarray}
{\mathcal{L}}_{a}&=&n_{a}\frac{\partial}{\partial k}- \frac{1}{2}{\rm ctg} \frac{k}{2}\varepsilon_{abc}n^{b} {\mathcal{A}}^{c}+ \frac{1}{2}{\mathcal{A}}_{a},\label{eq_3.36a} \\
{\mathcal{R}}_{a}&=&n_{a}\frac{\partial}{\partial k}- \frac{1}{2}{\rm ctg} \frac{k}{2}\varepsilon_{abc}n^{b} {\mathcal{A}}^{c}- \frac{1}{2}{\mathcal{A}}_{a},\label{eq_3.36b} \\
{\mathcal{L}}_{a}&=&n^{a}dk + 2 \sin^{2}\frac{k}{2}\varepsilon^{a}{}_{bc}n^{b}dn^{c}+ \sin k dn^{a},\label{eq_3.36c} \\
{\mathcal{R}}_{a}&=&n^{a}dk - 2 \sin^{2}\frac{k}{2}\varepsilon^{a}{}_{bc}n^{b}dn^{c}+ \sin k dn^{a}.\label{eq_3.36d}
\end{eqnarray}

Using the $\mathbb{R}^{3}$-vector notation, including also the vectors with operator components, we can denote briefly, without using indices and labels,
\begin{eqnarray}
\overline{{\mathcal{L}}}&=&\overline{n}\frac{\partial}{\partial k}- 
\frac{1}{2}{\rm ctg} \frac{k}{2}\overline{n} \times \overline{{\mathcal{A}}}+ \frac{1}{2}\overline{{\mathcal{A}}},\label{eq_3.37a} \\
\overline{{\mathcal{R}}}&=&\overline{n}\frac{\partial}{\partial k}- \frac{1}{2}{\rm ctg} \frac{k}{2}\overline{n} \times \overline{{\mathcal{A}}}- \frac{1}{2}\overline{{\mathcal{A}}},\label{eq_3.37b} \\
\underline{{\mathcal{L}}}&=&\overline{n}dk + 2 \sin^{2}\frac{k}{2}\overline{n} \times d\overline{n}+ \sin k d\overline{n},\label{eq_3.37c} \\
\underline{{\mathcal{R}}}&=&\overline{n}dk - 2 \sin^{2}\frac{k}{2}\overline{n} \times d\overline{n}+ \sin k d\overline{n},\label{eq_3.37d}\\
\overline{{\mathcal{A}}}&=& \overline{k}\times\overline{\nabla},\label{eq_3.37e}
\end{eqnarray}
where $\overline{\nabla}$ denotes the Euclidean gradient operator.

Let us note the following interesting and suggestive duality relations:
\begin{eqnarray}
\left\langle dk, {\mathcal{A}}_{a} \right\rangle = {\mathcal{A}}_{a}k = 0, &\quad& \left\langle dk, \frac{\partial}{\partial k} \right\rangle =1,\label{eq_3.38a} \\
\left\langle dn_{a}, {\mathcal{A}}_{b} \right\rangle = {\mathcal{A}}_{b}n_{a}= \varepsilon_{abc}n^{c}, &\quad& \left\langle dn_{a}, \frac{\partial}{\partial k} \right\rangle = \frac{\partial n_{a}}{\partial k}= 0.\label{eq_3.38b}
\end{eqnarray}

Obviously, in $\mathbb{R}^{3}$, considered as an Abelian group under addition of vectors, the right-invariant fields coincide with the left-invariant ones, and when using spherical variables we have then
\begin{eqnarray}
\overline{{\mathcal{L}}}=\overline{{\mathcal{R}}}&=&\overline{{\mathcal{\nabla}}}=\overline{n}\frac{\partial}{\partial r} - \frac{1}{r}\overline{n} \times \overline{{\mathcal{A}}}(\overline{r}),\label{eq_3.39a} \\
\underline{{\mathcal{L}}}=\underline{{\mathcal{R}}}&=&d\overline{r}= \overline{n}dr + rd\overline{n}.\label{eq_3.39b}
\end{eqnarray}
This is in agreement with the formulas (\ref{eq_3.37a})--(\ref{eq_3.37e}), namely, in a small neighbourhood of the group identity $I_{2} \in{\rm SU}(2)$, i.e., for $\overline{k} \approx \overline{0}$, expressions (\ref{eq_3.37a})--(\ref{eq_3.37e}) up to higher-order terms in $\overline{k}$, one obtains
\begin{eqnarray}
\overline{{\mathcal{L}}} \approx \overline{{\mathcal{R}}} &\approx& \overline{{\mathcal{\nabla}}}_{\overline{k}}=\overline{n}\frac{\partial}{\partial k} - \frac{1}{k}\overline{n} \times \overline{{\mathcal{A}}},\label{eq_3.40a} \\
\underline{{\mathcal{L}}} \approx \underline{{\mathcal{R}}} &\approx& d\overline{k} = \overline{n}dk + kd\overline{n}.\label{eq_3.40b}
\end{eqnarray}
Obviously, the quantities $\overline{n}$, $d\overline{n}$, $\overline{{\mathcal{A}}}$ are non-sensitive to the asymptotics $k \rightarrow 0$, because they are purely angular $(\theta, \varphi)$ variables, independent of $k$. 

${\rm SU}(2)$ is the sphere $S^{3}(0,1)$ in $\mathbb{R}^{4}$. Taking the sphere of radius $R$, $S^{3}(0,R) \subset \mathbb{R}^{4}$, and performing the limit transition $R \rightarrow \infty$, one obtains also the relationships (\ref{eq_3.39a}), (\ref{eq_3.39b}) as an asymptotic limit.

The Killing metric tensor with the modified normalization (\ref{eq_3.7}) is given by
\begin{equation}
g_{ij}= \frac{4}{k^{2}} \sin^{2}\frac{k}{2}\delta_{ij} + \left(1- \frac{4}{k^{2}} \sin^{2}\frac{k}{2}\right)\frac{k^{i}}{k}\frac{k^{j}}{k} \label{eq_3.41}
\end{equation}
and its contravariant inverse by
\begin{equation}
g^{ij}= \frac{k^{2}}{4\sin^{2}\frac{k}{2}} \delta^{ij} + \left(1- \frac{k^{2}}{4\sin^{2}\frac{k}{2}} \right)\frac{k^{i}}{k}\frac{k^{j}}{k}. \label{eq_3.42}
\end{equation}
Obviously, the corresponding metric element may be concisely written as 
\begin{equation}
ds^{2}=dk^{2}+ 4\sin^{2}\frac{k}{2}\left(d\theta^{2} + \sin^{2}\theta d\varphi^{2}\right) = dk^{2}+ 4\sin^{2}\frac{k}{2} d\overline{n} \cdot d\overline{n} \label{eq_3.43}
\end{equation}
or, in a more sophisticated way,
\begin{equation}
g= dk \otimes dk + 4\sin^{2}\frac{k}{2}\delta_{AB} dn^{A} \otimes dn^{A}, \label{eq_3.44}
\end{equation}
and, similarly, for the inverse tensor,
\begin{equation}
g^{-1}=\frac{\partial}{\partial k} \otimes \frac{\partial}{\partial k}+ \frac{1}{4\sin^{2}\frac{k}{2}}\delta^{AB}{\mathcal{A}}_{A} \otimes {\mathcal{A}}_{B}. \label{eq_3.45}
\end{equation}

According to the standard procedure, the volume element on the Riemannian manifold is given by
\begin{equation}
d\mu\left(\overline{k}\right)= \sqrt{\left|g\right|}d_{3}\overline{k}= 
\sqrt{{\rm det} \left[g_{ij}\left(\overline{k}\right)\right]}d_{3}\overline{k}. \label{eq_3.46}
\end{equation}

It is easy to see that for our normalization of the metric tensor,
\begin{equation}
d\mu\left(\overline{k}\right)= 4\sin^{2}\frac{k}{2} \sin \theta  dk 
d\theta d\varphi = \frac{4\sin^{2}\frac{k}{2}}{k^{2}}\ d_{3}\overline{k},\label{eq_3.47}
\end{equation}
where $d_{3}\overline{k}$ is the usual volume element in $\mathbb{R}^{3}$ as the space of rotation vectors $\overline{k}$. This volume element is identical with that given by (\ref{eq_3.47}), (\ref{eq_3.48}), (\ref{eq_3.49}). The reason is that all these expressions are translationally-invariant and the Haar measure is unique. We assume here that $G$ is unimodular; in fact, we mean only the compact semisimple groups and their products with Abelian groups (obviously in the latter case it is not the Killing tensor that is meant in the Abelian factor); nevertheless, any metric meant there is also assumed translationally-invariant, and so the total Riemann measure also coincides with (\ref{eq_3.47}), (\ref{eq_3.48}), (\ref{eq_3.49}).

Let us mention that when the Euler angles $(\varphi, \vartheta, \psi)$ are used as a parametri\-zation, then the Riemann metric is given by
\begin{equation}
ds^{2}= d\vartheta^{2}+ d\varphi^{2}+ 2 \cos \vartheta d\varphi d\psi + d\psi^{2}. \label{eq_3.48}
\end{equation}
The measure element is then expressed
\begin{equation}
d\mu\left(\varphi, \vartheta, \psi\right)= \sin \vartheta d\vartheta d\varphi d\psi. \label{eq_3.49}
\end{equation}
The metric element expression (\ref{eq_3.48}) may be diagonalized by introducing the new "angles":
\begin{equation}
\alpha = \varphi + \psi, \qquad \beta = \varphi - \psi, \label{eq_3.50}
\end{equation}
however, this representation rather is not used practically.

Let us remind that on ${\rm SU}(2)$ the range of Euler angles is $[0,4\pi]$ for $\varphi$, $\psi$, and $[0,2\pi]$ for $\vartheta$, on ${\rm SO}(3, \mathbb{R})$, it is respectively $[0,2\pi]$ and $[0,\pi]$.

In some of earlier formulas we used the convention of the Haar measure on compact groups normalized to unity, $\mu(G)=1$. When normalized in this way, it will be denoted as $\mu_{1}$. The label "(1)" will be omitted when the normalization is clear from the context or when there is no danger of confusion. 

After elementary integrations we find that on ${\rm SU}(2)$ the element of normalized measure is given by
\begin{equation}\label{eq_3.51}
d\mu_{(1)}=\frac{1}{4\pi^{2}}\sin^{2}\frac{k}{2}\sin\theta dk d\theta d\varphi=\frac{\sin^{2}\frac{k}{2}}{4\pi^{2}k^{2}}\ d_{3}\overline{k}.
\end{equation}
If we used the normalization (\ref{eq_3.47}), the "volume" of ${\rm SU}(2)$ would be $16\pi^{2}$. With the same normalization, the volume of ${\rm SO}(3,\mathbb{R})$ would be $8\pi^{2}$. It is intuitively clear: ${\rm SU}(2)$ is "twice larger" than ${\rm SO}(3,\mathbb{R})$. So, we would have
\begin{equation}\label{eq_3.52}
d\mu_{(1){\rm SO}(3,\mathbb{R})}=\frac{1}{2\pi^{2}}\sin^{2}\frac{k}{2}\sin\theta dk d\theta d\varphi=\frac{\sin^{2}\frac{k}{2}}{2\pi^{2}k^{2}}d_{3}\overline{k}.
\end{equation} 
However, as mentioned, all formulas will be meant in the covering group sense ${\rm SU}(2)$.

The metric tensor (\ref{eq_3.41}), (\ref{eq_3.43}) is conformally flat. It is seen when we introduce some new variables $\overline{\varrho}$ instead of $\overline{k}$, namely,
\begin{equation}\label{eq_3.53}
\varrho=|\overline{\varrho}|=a{\rm tg} \frac{k}{4}, \qquad \frac{\overline{\varrho}}{\varrho}=\frac{\overline{k}}{k}=\overline{n},
\end{equation}
where $a$ denotes some positive constant. Then (\ref{eq_3.43}) becomes 
\begin{equation}\label{eq_3.54}
ds^{2}=\frac{16a^{2}}{(a^{2}+\varrho^{2})^{2}}\left(d\varrho^{2}+
\varrho^{2}\left(d\theta^{2}+\sin^{2}\theta d\varphi^{2}\right)\right)=
\frac{16a^{2}}{(a^{2}+\varrho^{2})^{2}}\left(d\varrho^{2}+
\varrho^{2}d\overline{n}\cdot d\overline{n}\right),
\end{equation}
or, using again the "sophisticated" form (\ref{eq_3.44}),
\begin{equation}\label{eq_3.55}
g=\frac{16a^{2}}{(a^{2}+\varrho^{2})^{2}}\left(d\varrho\otimes d\varrho+\varrho^{2}\delta_{ab}dn^{a}\otimes dn^{b}\right).
\end{equation}
Obviously, (\ref{eq_3.53}) is a conformal mapping of ${\rm SU}(2)$ onto $\mathbb{R}^{3}$ with its usual Euclidean metric. 
The ball $K^{2}(0,2\pi)$ "blows up" to the total $\mathbb{R}^{3}$ and the sphere $S^{2}(0,2\pi)$ "blows up" to infinity. In other words, ${\rm SU}(2)$ is identified with the one-point compactification of $\mathbb{R}^{3}$
and the element $-I_{2}\in {\rm SU}(2)$ becomes just the compactifying point. The ball $K^{2}(0,\pi)$ corresponding
to the manifold of ${\rm SO}(3,\mathbb{R})$ and its boundary sphere  $S^{2}(0,\pi)$ (non-trivial square-root of identity) become respectively $K^{2}(0,a)$ and $S^{2}(0,a)$. If we put $a=\pi$, they are mapped onto themselves.
From the conformal point of view the particular choice of the constant $a$ does not matter.

The projective mapping (\ref{eq_3.17a}) of ${\rm SO}(3, \mathbb{R})$ onto ${\rm P}\mathbb{R}^{3}$ maps geodetics of (\ref{eq_3.43}) onto straight lines in $\mathbb{R}^{3}$. However, it is neither isometry nor the conformal transformation, instead we have that
\begin{equation}\label{eq_3.56}
ds^{2}=\frac{4}{4+\varkappa^{2}}\left(\frac{4}{4+\varkappa^{2}}d\varkappa^{2}+
\varkappa^{2}\left(d\theta^{2}+\sin^{2}\theta d\varphi^{2}\right)\right).
\end{equation}

On ${\rm SU}(2)$ the formulas (36), (41) in \cite{2} take on the following form: 
\begin{equation}\label{eq_3.57}
\mathcal{L}^{a}(u)=R(u)^{a}{}_{b}\mathcal{R}^{b}(u), \qquad \mathcal{L}_{a}(u)=\mathcal{R}_{b}(u)R(u)^{-1b}{}_{a},
\end{equation}
where the dependence ${\rm SU}(2)\ni u\mapsto R(u)\in {\rm SO}(3, \mathbb{R})$ is given by (\ref{eq_3.5}), (\ref{eq_3.23}). Obviously, in orthonormal coordinates this is the same formula, because inverses of orthogonal matrices coincide with their transposes. The corresponding symmetric operators $\mathbf{\Sigma}_{a}$, $\widehat{\mathbf{\Sigma}}_{a}$, denoted respectively by
\begin{equation}\label{eq_3.58}
\mathbf{S}_{a}=\frac{\hbar}{i}\mathcal{L}_{a}, \qquad \widehat{\mathbf{S}}_{a}=\frac{\hbar}{i}\mathcal{R}_{a},
\end{equation}
are, obviously, interrelated by the same formula,
\begin{equation}\label{eq_3.59}
\mathbf{S}_{a}(u)=\widehat{\mathbf{S}}_{b}(u)R(u)^{-1b}{}_{a}.
\end{equation}
When interpreted in terms of action on the wave functions on ${\rm SU}(2)$, they are operators of rotational angular momentum (spin) respectively in the spatial and co-moving representations. The corresponding operators of hyperspin 
\begin{equation}\label{eq_3.60a}
\mathbf{\Delta}_{a}=\frac{\hbar}{i}\mathbf{\mathcal{A}}_{a}=
\frac{\hbar}{i}\varepsilon_{ab}{}^{c}k^{b}\frac{\partial}{\partial k^{c}} 
\end{equation}
are given by
\begin{equation}\label{eq_3.60}
\mathbf{\Delta}_{a}=\mathbf{S}_{a}-\widehat{\mathbf{S}}_{a}, \qquad \mathcal{A}_{a}=\mathcal{L}_{a}-\mathcal{R}_{a}.
\end{equation}
The term "hyper" is used because this quantity tells us "how much" the spatial components of spin exceed the corresponding laboratory ones. The operators $\mathcal{A}_{a}$ generate rotations of the rotation vector; just the meaning of "hyper".

According to (30) in \cite{2}, the corresponding classical quantities are given by 
\begin{eqnarray}
&&S_{a}=p_{j}\mathcal{L}^{j}{}_{a}, \qquad \widehat{S}_{a}=p_{j}\mathcal{R}^{j}{}_{a}=S_{b}R^{b}{}_{a},\label{eq_3.61a}\\
&&\Delta_{a}=p_{j}\Delta^{j}{}_{a}=S_{a}-\widehat{S}_{a}=
\varepsilon_{ab}{}^{c}k^{b}p_{c},\label{eq_3.61b}
\end{eqnarray}
where $p_{j}$ denote canonical momenta conjugate to $k^{j}$ or rather to the corresponding generalized velocities $dk^{j}/dt$. 

Evaluating differential forms on vector tangent to trajectories in the configuration spaces  ${\rm SU}(2)$, ${\rm SO}(3, \mathbb{R})$, we obtain the following quantities:
\begin{equation}\label{eq_3.62}
\omega^{a}=\mathcal{L}^{a}{}_{i}\left(\overline{k}\right)\frac{dk^{i}}{dt}, \qquad \widehat{\omega}^{a}=\mathcal{R}^{a}{}_{i}\left(\overline{k}\right)
\frac{dk^{i}}{dt}, \qquad \omega^{a}\left(u,\dot{u}\right)=
R(u)^{a}{}_{b}\widehat{\omega}^{a}\left(u,\dot{u}\right).
\end{equation}
They are respectively spatial $(\omega^{a})$ and co-moving $(\widehat{\omega}^{a})$ components of angular velocity. They are non-holonomic, i.e., fail to be time derivatives of any generalized coordinates. The following duality relations are satisfied:
\begin{equation}\label{eq_3.63}
s_{a}\omega^{a}=\widehat{s}_{a}\widehat{\omega}^{a}=p_{i}\frac{dk^{i}}{dt}.
\end{equation}

Let us quote the obvious commutators and Poisson brackets:
\begin{eqnarray}
[\mathcal{L}_{a}, \mathcal{L}_{b}]=-\varepsilon_{ab}{}^{c}\mathcal{L}_{c}, \quad 
[\mathcal{R}_{a}, \mathcal{R}_{b}]=\varepsilon_{ab}{}^{c}\mathcal{R}_{c}, \quad 
[\mathcal{L}_{a}, \mathcal{R}_{b}]=0,\label{eq_3.64a}\\
\left[\mathcal{A}_{a}, \mathcal{L}_{b}\right]=-\varepsilon_{ab}{}^{c}\mathcal{L}_{c}, \quad 
[\mathcal{A}_{a}, \mathcal{R}_{b}]=-\varepsilon_{ab}{}^{c}\mathcal{R}_{c}, \quad 
[\mathcal{A}_{a}, \mathcal{A}_{b}]=-\varepsilon_{ab}{}^{c}\mathcal{A}_{c},
\label{eq_3.64b}\\
\frac{1}{\hbar i}[\mathbf{S}_{a}, \mathbf{S}_{b}]=\varepsilon_{ab}{}^{c}\mathbf{S}_{c}, \quad 
\frac{1}{\hbar i}[\widehat{\mathbf{S}}_{a},\widehat{\mathbf{S}}_{b}]=
-\varepsilon_{ab}{}^{c}\widehat{\mathbf{S}}_{c}, \quad 
\frac{1}{\hbar i}[\mathbf{S}_{a},\widehat{\mathbf{S}}_{b}]=0,\label{eq_3.65a}\\
\frac{1}{\hbar i}[\mathbf{\Delta}_{a}, \mathbf{S}_{b}]=\varepsilon_{ab}{}^{c}\mathbf{S}_{c}, \quad \frac{1}{\hbar i}[\mathbf{\Delta}_{a},\widehat{\mathbf{S}}_{b}]=
\varepsilon_{ab}{}^{c}\widehat{\mathbf{S}}_{c}, \quad 
\frac{1}{\hbar i}[\mathbf{\Delta}_{a},\mathbf{\Delta}_{b}]=
-\varepsilon_{ab}{}^{c}\mathbf{\Delta}_{c},\label{eq_3.65b}\\
\{S_{a}, S_{b}\}=\varepsilon_{ab}{}^{c}S_{c}, \quad 
\{\widehat{S}_{a},\widehat{S}_{b}\}=-\varepsilon_{ab}{}^{c}\widehat{S}_{c},\quad 
\{S_{a},\widehat{S}_{b}\}=0,\label{eq_3.66a}\\
\{\Delta_{a}, S_{b}\}=\varepsilon_{ab}{}^{c}S_{c}, \quad 
\{\Delta_{a}, \widehat{S}_{b}\}=\varepsilon_{ab}{}^{c}\widehat{S}_{c},\quad \{\Delta_{a}, \Delta_{b}\}=-\varepsilon_{ab}{}^{c}\Delta_{c}.\label{eq_3.66b}
\end{eqnarray}

In the enveloping algebras built over Lie algebras of $\mathcal{L}$- and $\mathcal{R}$-operators there exists only one Casimir invariant, namely, the second-order one,
\begin{equation}\label{eq_3.67}
C(\mathcal{L}, 2)=C(\mathcal{R}, 2)=\Delta=\delta^{ab}\mathcal{L}_{a}\mathcal{L}_{b}=
\delta^{ab}\mathcal{R}_{a}\mathcal{R}_{b}.
\end{equation}

In physical expressions like various kinetic energies and so on, one uses their $(-\hbar^{2})$-multiplies,
\begin{equation}\label{eq_3.68}
\mathbf{S}^{2}:=C(S, 2)=C\left(\widehat{S}, 2\right)=-\hbar^{2}\Delta.
\end{equation}

Obviously, there is also only one Casimir in the associative algebra generated by the Lie algebra of $\mathcal{A}$-operators,
\begin{equation}\label{eq_3.69}
\mathcal{A}^{2}:=C(\mathcal{A}, 2)=\delta^{ab}\mathcal{A}_{a}\mathcal{A}_{b}, \qquad \Delta^{2}:=-\hbar^{2}\mathcal{A}^{2}.
\end{equation}
After some easy calculations one obtains for (\ref{eq_3.67})
\begin{equation}\label{eq_3.70}
\Delta=\frac{\partial^{2}}{\partial k^{2}}+{\rm ctg}\frac{k}{2}\frac{\partial}{\partial k}+\frac{1}{4\sin^{2}k/ 2}A^{2}.
\end{equation}

For obvious reasons, when expressed by the spherical angular variables $(\theta, \varphi)$ in the space of rotation vector $\overline{k}$, $\Delta^{2}$ has identical form with the operator of the squared magnitude of orbital angular momentum. Its spectrum consists obviously of non-negative numbers $\hbar^{2} l(l+1)$, where $l$ denotes non-negative integers, $l=0, 1, 2, \dots $. As we saw in (106) in \cite{2}, $\Delta^{2}=-\hbar^{2}\mathcal{A}^{2}$ does commute with the Laplace-Beltrami Casimir $S^{2}=-\hbar^{2}\delta^{ab}\mathcal{L}_{a}\mathcal{L}_{b}=
-\hbar^{2}\delta^{ab}\mathcal{R}_{a}\mathcal{R}_{b}$, so they have common wave functions. Spectrum of the Laplace-Beltrami operator consists of non-negative numbers $\hbar^{2}j(j+1)$, where $j$ runs over non-negative half-integers and integers, $j=0, \frac{1}{2}, 1, \frac{3}{2}, \dots$, i.e., $j\in \{0\} \bigcup \left(\mathbb{N}/2\right)$. Obviously, $\mathbb{N}$ denotes the set of naturals and $j$ is just the label of irreducible unitary representations of ${\rm SU}(2)$. When $j$ is fixed, then $l$ runs over the range $l=0, 1,\ldots, 2j$ for the possible common eigenfunctions of $S^{2}$ and $\Delta^{2}$. According to (97)--(102), (109), (110) in \cite{2}, we have that
\begin{equation}\label{eq_3.71}
\mathbf{S}^{2}D(j)=\hbar^{2}j(j+1)D(j), \qquad \mathbf{S}^{2}\varepsilon(j)=\hbar^{2}j(j+1)\varepsilon(j),
\end{equation}
i.e., all matrix elements of the $j$-th irreducible unitary representation, or, equivalently, all elements of the minimal two-sided ideal $M(j)$, are eigenfunctions of
\begin{equation}\label{eq_3.72}
\mathbf{S}^{2}=\delta^{ab}\mathbf{S}_{a}\mathbf{S}_{b}=
\delta^{ab}\widehat{\mathbf{S}}_{a}\widehat{\mathbf{S}}_{b}=-\hbar^{2}\Delta[g]
\end{equation}
with eigenvalues $\hbar^{2}j(j+1)$.

Further, we have the following algebraization of operators
\begin{equation}\label{eq_3.72a}
\mathbf{S}_{a}=\frac{\hbar}{i}\mathcal{L}_{a},\qquad \widehat{\mathbf{S}}_{a}=\frac{\hbar}{i}\mathcal{R}_{a}
\end{equation}
in this representation:
\begin{eqnarray}
&&S_{a}D(j)=S(j)_{a}D(j), \qquad \widehat{S}_{a}D(j)=D(j)S(j)_{a},\label{eq_3.73a}\\ 
&&\Delta_{a}D(j)=[S(j)_{a}, D(j)],\label{eq_3.73b}
\end{eqnarray}
and similarly for elements of the canonical basis, because, as we saw, there is a proportionality:
\begin{equation}\label{eq_3.74}
\varepsilon(j)_{km}=(2j+1)D(j)_{km}.
\end{equation}
Here $S(j)_{a}$ are standard $(2j+1)\times (2j+1)$ Hermitian matrices of the $j$-labelled angular momentum. According to 
(89), (92) in \cite{2} we have that
\begin{equation}\label{eq_3.75}
D(j)\left(u\left(\overline{k}\right)\right)=\exp \left(\frac{i}{\hbar} k^{a}S(j)_{a}\right). 
\end{equation}
This algebraization of differential operators is very convenient because the matrices of angular momentum are standard. Therefore, (114)--(117) in \cite{2}, or, alternatively, (118)--(121) in \cite{2}, may be used; obviously the label $\alpha$ to be replaced by $j$ and the symbols $\Sigma(\alpha)_{a}$ by $S(j)_{a}$.

Representations $D(j)$ are irreducible, so, by definition,
\begin{equation}\label{eq_3.76}
\delta^{ab}S(j)_{a}S(j)_{b}=\sum_{a}S(j)^{2}{}_{a}=
\hbar^{2}j(j+1){\rm Id}_{(2j+1)}.
\end{equation}

The only Abelian Lie subgroups of ${\rm SU}(2)$, ${\rm SO}(3, \mathbb{R})$ are one-dimensional, just the one-parameter subgroups. Therefore, one can choose only one $\mathcal{L}$-type operator and only one $\mathcal{R}$-type operator to form, together with $-\hbar^{2}\Delta[g]$, the complete system of eigenequations for the functions $\varepsilon(j)_{kl}/D(j)_{kl}$. Traditionally one chooses for $S(j)_{a}$ such a representation that $S(j)_{3}$ are diagonal. Then, of course, one should choose the operators $\mathcal{L}_{3}$, $\mathcal{R}_{3}$, or in terms of observables $\mathbf{S}_{3}$, $\widehat{\mathbf{S}}_{3}$. This is obviously, the matter of convention; one could as well take any versor $\overline{n}\in \mathbb{R}^{3}$ and operators $n^{a}\mathcal{L}_{a}$, $n^{a}\mathcal{R}_{a}$ (or $n^{a}\mathbf{S}_{a}$, $n^{a}\widehat{\mathbf{S}}_{a}$), assuming only that $n^{a}S(j)_{a}$ is diagonal for any $j$. When we fix the quantum number $j$, then the eigenvalues of $\mathbf{S}_{3}$, $\widehat{\mathbf{S}}_{3}$ have the form $\hbar m$, where $m=-j, -j+1, \dots, j-1, j$, jumping by one. Therefore, the matrix labels of $D(j)_{mk}$, $\varepsilon(j)_{mk}$ are not taken as $1, \dots, 2j+1$, but rather as $-j, -j+1, \dots, j-1, j$. The matrices $S(j)_{3}$ are then chosen as
\begin{eqnarray}\label{eq_3.77}
S(j)_{3}&=&{\rm Diag}(-\hbar j, -\hbar(j-1), \dots, \hbar(j-1), \hbar j)\nonumber\\
&=&\hbar {\rm Diag}(-j, -(j-1), \dots, (j-1), j).
\end{eqnarray}

Therefore, the basic functions $D(j)_{mk}$, $\varepsilon(j)_{mk}=(2j+1)D(j)_{mk}$ are defined by the following maximal system of compatible eigenequations:
\begin{eqnarray}
\mathbf{S}^{2}D(j)_{mk}&=&\hbar^{2}j(j+1)D(j)_{mk},\label{eq_3.78a}\\
\mathbf{S}_{3}D(j)_{mk}&=&m\hbar D(j)_{mk},\label{eq_3.78b}\\
\widehat{\mathbf{S}}_{3}D(j)_{mk}&=&k\hbar D(j)_{mk}.\label{eq_3.78c}
\end{eqnarray}
The solution is unique up to normalization and this one is fixed by the first and third equations in (\ref{eq_3.73a}), (\ref{eq_3.73b}) with $n(\alpha)=n(j)=2j+1$.

Quite independently on the representation theory, the functions $D(j)_{mk}$ as solutions of (\ref{eq_3.78a})--(\ref{eq_3.78c}) were found as basic stationary states of the free symmetric top, i.e., one with the following Hamiltonian (kinetic energy):
\begin{equation}\label{eq_3.79}
\mathbf{H}=\frac{1}{2I}\left(\widehat{\mathbf{S}}_{1}\right)^{2}+
\frac{1}{2I}\left(\widehat{\mathbf{S}}_{2}\right)^{2}+
\frac{1}{2K}\left(\widehat{\mathbf{S}}_{3}\right)^{2}.
\end{equation}
The corresponding energy levels (eigenvalues of energy) are given by
\begin{equation}\label{eq_3.80}
E_{j,k}=\frac{1}{2I}\hbar^{2}j(j+1)+\left(\frac{1}{2K}-
\frac{1}{2I}\right)\hbar^{2}k^{2}.
\end{equation}
Obviously, they are $2(2j+1)$-fold degenerate, i,e, they do not depend on $m$ at all and they do not distinguish the sign of $k$. If the top is spherical, $K=I$, they are, obviously, $(2j+1)^{2}$-fold degenerate. When the top is completely asymmetric, the energy levels are $(2j+1)$-fold degenerate (independence on the spatial quantum number $m$).

Matrix elements $D(j)_{mk}$ of irreducible unitary representations, i.e., equivalently, elements of the canonical basis $\varepsilon(j)_{mk}=(2j+1)D(j)_{mk}$ are common solutions of the system of eigenequations (\ref{eq_3.78a})--(\ref{eq_3.78c}).

There is also another complete system of commuting operators, namely, $\mathbf{S}^{2}$, $\mathbf{\Delta}^{2}$, $\mathbf{\Delta}_{3}$; obviously, taking the third component is but just a custom, we could take as well $n^{a}\mathbf{\Delta}_{a}$, where $\overline{n}$ is an arbitrary unit vector in $\mathbb{R}^{3}$. Obviously, any common eigenfunction of $ \Delta^{2}, \Delta_{3}$ has the following form:
\begin{equation}\label{eq_3.81}
\psi\left(\overline{k}\right)=\psi\left(k, \theta, \varphi\right)=f(k)Y_{lm}\left(\overline{n}\left(\theta, \varphi\right)\right),
\end{equation}
where $f$ is an arbitrary function of the "rotation angle" $k=\left|\overline{k}\right|$ and $\overline{n}$ is the unit vector of the oriented rotation axis; $Y_{lm}$ is the standard symbol of spherical functions. The eigenvalues are respectively given by $\hbar^{2}l(l+1)$, where $l\in \{0\}\bigcup \mathbb{N}$ is an arbitrary non-negative integer, and $m\hbar$, where $m$ runs over the range $m=-l, -l+1, \dots, l-1, l$, jumping by one. The well-known system of eigenequations is satisfied:
\begin{equation}\label{eq_3.82}
\Delta^{2}\psi=\hbar^{2} l(l+1)\psi, \quad \Delta_{3}\psi=\hbar m \psi.
\end{equation}

The function $f$ is arbitrary, because it is transparent for the action of $\Delta^{2}$, $\Delta_{3}$. The space of solutions of (\ref{eq_3.82}) is infinite-dimensional; this infinity is due to the arbitrariness of $f$. Roughly speaking, for any fixed value of $l$, such a system of functions represents an irreducible tensor of the group of automorphisms (\ref{eq_3.23}). The value $l=0$ corresponds to scalars, i.e., functions constant on classes of adjoint elements. They are linear combinations or rather series of idempotents/characters $\varepsilon(j)/\chi(j)$. Similarly, all higher-order tensors may be combined from their orthogonal projections onto ideals $M(j)$. Those projections are common eigenfunctions
\begin{equation}\label{eq_3.82a}
Q\{j\}_{lm}=f_{jl}(k)Y^{l}{}_{m}\left(\overline{n}\right)
\end{equation}
of $\mathbf{S}^{2}$, $\mathbf{\Delta}^{2}$, $\mathbf{\Delta}_{3}$, therefore, the "radial" functions $f_{jl}$ satisfy the following reduced eigenequation:
\begin{equation}\label{eq_3.83}
\frac{d^{2}f_{jl}}{dk^{2}}+{\rm ctg}\frac{k}{2}\frac{df_{jl}}{dk}+\left( j(j+1)-\frac{l(l+1)}{4\sin^{2}k/2}\right)f_{jl}=0.
\end{equation}
When $j$ is fixed, then $l$ runs over the range
\begin{equation}\label{eq_3.84}
l=0, 1, \dots, 2j-1, 2j,
\end{equation}
i.e., integers from $0$ to $2j$. It turn, any $l$-level is $(2j+1)$-fold degenerate, thus, for any fixed $j$, the number of independent functions $Q\{j\}_{lm}$ equals
\begin{equation}\label{eq_3.85}
\sum^{2j}_{l=0}(2l+1)=(2j+1)^{2},
\end{equation}
just as expected, because $\dim M(j)=(2j+1)^{2}$.

This is an alternative choice of basis, or rather of orthonormal complete system in $L^{2}\left({\rm SU}(2)\right)$, tensorially ruled by irreducible representations of ${\rm SO}(3,\mathbb{R})$ as the automorphism group of ${\rm SU}(2)$.

The corresponding finite transformation rule reads
\begin{eqnarray}\label{eq_3.86}
Q\{j\}_{lm}\left(gu\left(\overline{k}\right)g^{-1}\right)&=&
Q\{j\}_{lm}\left(u\left(R(g)\right)\overline{k}\right)=
Q\{j\}_{lm}\left(k,R(g)\overline{n}\right)\nonumber\\
&=&\sum_{n} Q\{j\}_{ln}\left(k,\overline{n}\right)D(l)_{nm}\left(R(g)\right).
\end{eqnarray}
Infinitesimally this is expressed as
\begin{equation}\label{eq_3.87}
\Delta_{a}Q\{j\}_{lm}=\sum_{n} Q\{j\}_{ln}S(l)_{anm}.
\end{equation}
In terms of the convolution commutator:
\begin{equation}\label{eq_3.88}
\left[\frac{\hbar}{i}\mathcal{L}_{a}\delta,Q\{j\}_{lm}\right]=
\left[\frac{\hbar}{i}\mathcal{L}_{a}\varepsilon(j),Q\{j\}_{lm}\right]=
\sum_{n} Q\{j\}_{ln}S(l)_{anm}.
\end{equation}
Obviously, the convolution commutator is meant in the sense
\begin{equation}\label{eq_3.89}
[f,g]=f\ast g-g\ast f.
\end{equation}
The use of spherical functions $Y^{l}{}_{m}\left(\overline{n}\right)$ in (\ref{eq_3.82}) expresses explicitly the fact that for a fixed $l$ we are dealing with an irreducible object of the group of inner automorphisms. This is so-to-speak a non-redundant description of such objects, with all its advantages and disadvantages. The obvious disadvantage is that the tensorial structure is hidden. The point is that $Y^{l}{}_{m}\left(\overline{n}\right)$ are independent quantities extracted from the $l$-th tensorial power of the unit versor $\overline{n}$, $\underset{l}{\otimes}\overline{n}$; analytically such a symmetric tensor is given by the system of components
\begin{equation}\label{eq_3.90}
n^{a_{1}}\ldots n^{a_{l}}.
\end{equation}
The transformation rule under $R\in{\rm SO}(3,\mathbb{R})$,
\begin{equation}\label{eq_3.91}
\left(R\overline{n}\right)^{a_{1}}\ldots\left(R\overline{n}\right)^{a_{l}}=
R^{a_{1}}{}_{b_{1}}\ldots R^{a_{l}}{}_{b_{l}}n^{b_{1}}\ldots n^{b_{l}}
\end{equation}
is evidently tensorial and preserves the symmetry, however, it is reducible, because orthogonal transformations preserve all trace operations. Irreducible objects are obtained from (\ref{eq_3.90}) by eliminating all traces, e.g., 
\begin{eqnarray}
\mathcal{Y}(1)^{a}&=&n^{a},\label{eq_3.92a}\\
\mathcal{Y}(2)^{ab}&=&n^{a}n^{b}-\frac{1}{3}\delta^{ab},\label{eq_3.92b}\\
\mathcal{Y}(3)^{abc}&=&n^{a}n^{b}n^{c}-\frac{1}{5}\left(n^{a}\delta^{bc}+
n^{b}\delta^{ca}+n^{c}\delta^{ab}\right),\label{eq_3.92c}\\
\mathcal{Y}(4)^{abcd}&=&n^{a}n^{b}n^{c}n^{d}-\frac{1}{7}\left(
n^{a}n^{b}\delta^{cd}+n^{a}n^{c}\delta^{bd}+n^{a}n^{d}\delta^{bc}+
n^{b}n^{c}\delta^{ad}\right.\nonumber\\
&+&n\left.^{b}n^{d}\delta^{ac}+n^{c}n^{d}\delta^{ab}\right)+
\frac{1}{35}\left(\delta^{ab}\delta^{cd}+\delta^{ac}\delta^{bd}+
\delta^{ad}\delta^{bc}\right),\label{eq_3.92d}
\end{eqnarray}
and so on. The logic of those tensors is that they are algebraically built of $n^{a}$, $\delta^{bc}$, are completely symmetric and traceless in any pair of indices (trace meant as a contraction with an appropriate $\delta_{ab}$).

Any $\mathcal{Y}(l)$ has only $(2l+1)$ independent components; they are linear combinations of $Y^{l}{}_{m}$, $m=-l,\ldots,l$. Therefore, the representation is very redundant, however, the tensorial structure is explicitly visible. Instead of functions $Q\{j\}_{lm}$ one can use tensorial objects
\begin{equation}\label{eq_3.93}
Q\{j,l\}^{a_{1}\ldots a_{l}}=f_{jl}(k)\mathcal{Y}(l)
\left(\overline{n}\right)^{a_{1}\ldots a_{l}}.
\end{equation}
Infinitesimally, the tensorial character of quantities $Q\{j,l\}$ is represented by the following relationship:
\begin{equation}\label{eq_3.94}
\mathcal{A}_{b}Q\{j,l\}^{a_{1}\ldots a_{l}}=
\left[\mathcal{L}_{b}\delta,Q\{j,l\}^{a_{1}\ldots a_{l}}\right]=
-\sum_{c}\varepsilon_{b}{}^{a_{c}}{}_{d}Q\{j,l\}^{a_{1}\ldots a_{c-1}da_{c+1}\ldots a_{l}},
\end{equation}
for example,
\begin{equation}\label{eq_3.95}
\mathcal{A}_{b}Q\{j,2\}^{km}=
\left[\mathcal{L}_{b}\delta,Q\{j,2\}^{km}\right]=
-\varepsilon_{b}{}^{k}{}_{d}Q\{j,2\}^{dm}
-\varepsilon_{b}{}^{m}{}_{d}Q\{j,2\}^{kd},
\end{equation}
and so on. Obviously, $\mathcal{L}_{b}\delta$ in these equations may be replaced by $\mathcal{R}_{b}\delta$ and both may be replaced by $\mathcal{L}_{a}\varepsilon(j)=\mathcal{R}_{a}\varepsilon(j)$. Irreducibility implies that
\begin{eqnarray}\label{eq_3.96}
\delta^{ab}\mathcal{A}_{a}\mathcal{A}_{b}Q\{j,l\}^{a_{1}\ldots a_{l}}&=&
\delta^{ab}\left[\mathcal{L}_{a}\delta,\left[\mathcal{L}_{b}\delta,
Q\{j,l\}^{a_{1}\ldots a_{l}}\right]\right]\nonumber\\
=\delta^{ab}\left[\mathcal{L}_{a}\varepsilon(j),
\left[\mathcal{L}_{b}\varepsilon(j),
Q\{j,l\}^{a_{1}\ldots a_{l}}\right]\right]&=&
-l(l+a)Q\{j,l\}^{a_{1}\ldots a_{l}}
\end{eqnarray}
with the (\ref{eq_3.89})-meaning of the convolution commutator.

There are some subtle points concerning irreducible tensors of the automorphism group, which were partially mentioned earlier in the paper devoted to general Lie groups \cite{2}. Namely, the tensorial quantities (131) in \cite{2} were introduced there. They were obtained as convolution monomials of $Q_{a}=\mathcal{L}_{a}\delta=\mathcal{R}_{a}\delta$ or $\Sigma_{a}=(\hbar/i)Q_{a}$, or rather as symmetrizations of these monomials. The symmetrizations of monomials built of $\Sigma_{a}$ are Hermitian in the sense of group algebra, just as $\Sigma_{a}$ themselves. However, in general they are not irreducible tensors, just the traces (in the sense of Killing metric tensor) must be subtracted. The symmetrized monomials are represented in the Peter-Weyl sense by matrices (130) in \cite{2} alternatively, depending on whether the convention (114) or (118) in \cite{2} is used. And an important point is of course that the monomials (122) in \cite{2} are different from the pointwise products $Q_{a}Q_{b}\ldots Q_{k}$. In particular, the pointwise products $Q(\alpha)_{a}Q(\alpha)_{b}\ldots Q(\alpha)_{k}$ of $M(\alpha)$-projections do not belong to $M(\alpha)$, whereas the convolutions $Q(\alpha)_{a}*Q(\alpha)_{b}*\cdots*Q(\alpha)_{k}$ obviously do.

Let us specialize the problem to ${\rm SU}(2)$. The distribution $\Sigma_{a}=(\hbar/i)Q_{a}$, physically corresponding to the angular momentum, is suggestively expressed by the operators (\ref{eq_3.58}),
\begin{equation}\label{eq_3.97}
\Sigma_{a}=\mathbf{S}_{a}\delta=\mathbf{\widehat{S}}_{a}\delta=
\frac{\hbar}{i}\mathcal{L}_{a}\delta=\frac{\hbar}{i}\mathcal{R}_{a}\delta,
\end{equation}
and its projections onto ideals $M(j)$ are given by
\begin{equation}\label{eq_3.98}
\Sigma(j)_{a}=\frac{\hbar}{i}\mathcal{L}_{a}\varepsilon(j)=
\frac{\hbar}{i}\mathcal{R}_{a}\varepsilon(j).
\end{equation}
Obviously, (127) in \cite{2} is a series built of (129) in \cite{2} with all possible values of $j=0,1/2,1,3/2,\ldots$; the limit being meant in the distribution sense. But of course $\Sigma(j)_{a}$ themselves are well-defined smooth functions and
\begin{equation}\label{eq_3.99}
\Sigma(j)_{a}=\frac{d\varepsilon(j)}{dk}\frac{k_{a}}{k}=
(2j+1)\frac{d\chi(j)}{dk}n_{a},
\end{equation}
because the idempotents $\varepsilon(j)$/characters $\chi(j)$ depend only on $k$. The Peter-Weyl coefficients of $\Sigma_{a}$ are given by the usual $(2j+1)\times(2j+1)$ matrices $S(j)_{a}$ of angular momentum or by their transposes $S(j)^{T}_{a}$, depending on which one of conventions (118), (114) in \cite{2} is used.

The higher-order Hermitian ${\rm SO}(3,\mathbb{R})$-tensors are again given by (131) in \cite{2} and the corresponding Peter-Weyl representing matrices (130) in \cite{2} will be denoted by
\begin{eqnarray}
S(j,l)_{a_{1}\ldots a_{l}}&=&
S(j)_{(a_{1}}\ldots S(j)_{a_{l})},\label{eq_3.100}\\
S(j,l)^{T}_{a_{1}\ldots a_{l}}&=&
S(j)^{T}_{(a_{1}}\ldots S(j)^{T}_{a_{l})}.\label{eq_3.101}
\end{eqnarray}
They are tensorial and symmetric, nevertheless, just like (\ref{eq_3.90}), they are still reducible. To obtain irreducible objects, one must eliminate traces, in analogy to (\ref{eq_3.92a})--(\ref{eq_3.92d}). The corresponding traceless parts of (matrix-valued) tensors $S(j,l)$, $S(j,l)^{T}$ will be denoted by
\begin{equation}\label{eq_3.102}
{}^{0}S(j,l)={\rm Traceless}\left(S(j,l)\right),\qquad
{}^{0}S(j,l)^{T}={\rm Traceless}\left(S(j,l)^{T}\right).
\end{equation}
Let us observe that the very literal analogy with (\ref{eq_3.92a})--(\ref{eq_3.92d}) is, nevertheless, misleading, because in (\ref{eq_3.82}), (\ref{eq_3.90}), (\ref{eq_3.92a})--(\ref{eq_3.92d}) we are dealing with the pointwise products
\begin{equation}\label{eq_3.103}
n^{a}n^{b}\ldots n^{r}\qquad {\rm or}\qquad k^{a}k^{b}\ldots k^{r}.
\end{equation}
Because of this the shape factor $f_{jl}(k)$ in (\ref{eq_3.82}), (\ref{eq_3.93}) must be introduced and subject to the "radial" Schr\"odinger-type equation. Unlike this, there is no problem of "radial" equation when one deals with functions $\Sigma(j,l)$ on ${\rm SU}(2)$ with the Peter-Weyl coefficients (\ref{eq_3.102}). Namely, for any fixed half/integer $j$ and any $l\leq 2j$, the following functions on ${\rm SU}(2)$
\begin{equation}\label{eq_3.104}
T(j,l)_{a_{1}\ldots a_{l}}={\rm Tr}\left({}^{0}S(j,l)_{a_{1}\ldots a_{l}}
\widehat{\varepsilon}(j)\right)={\rm Tr}\left({}^{0}S(j,l)_{a_{1}\ldots a_{l}}
D(j)(2j+1)\right)
\end{equation}
are eigenfunctions of $\mathbf{S}^{2}=-\hbar^{2}\mathbf{\Delta}$ with the eigenvalue $\hbar^{2}j(j+1)$, thus, they are elements of $M(j)$ and simultaneously are the eigenfunctions of $\mathbf{\Delta}^{2}=\delta^{ab}\mathbf{\Delta}_{a}\mathbf{\Delta}_{b}$ with the eigenvalue $\hbar^{2}l(l+1)$. Any element of $M(j)$ may be uniquely expanded as follows:
\begin{equation}\label{eq_3.105}
F=\sum^{2j}_{l=0}P(l)^{a_{1}\ldots a_{l}}T(j,l)_{a_{1}\ldots a_{l}},
\end{equation}
where the tensor $P(l)$ is totally symmetric and traceless.

Its Peter-Weyl matrix of coefficients $\widehat{F}$ in the convention (118) in \cite{2} has the following form:
\begin{equation}\label{eq_3.106}
\widehat{F}=\sum^{2j}_{l=0}P(l)^{a_{1}\ldots a_{l}}\ 
{}^{0}S(j,l)_{a_{1}\ldots a_{l}}.
\end{equation}
Obviously, the function (\ref{eq_3.105}) is Hermitian in the sense of group algebra if and only if the coefficients $P(l)^{a_{1}\ldots a_{l}}$ are real, because all the matrices ${}^{0}S(j,l)_{a_{1}\ldots a_{l}}$ combined in (\ref{eq_3.106}) are Hermitian.

The above representation is tensorially symmetric, however, informationally redundant. In non-redundant description, based on spherical functions $Y_{lm}$, we have instead of (\ref{eq_3.105}) the representation
\begin{equation}\label{eq_3.107}
F=\sum^{2j}_{l=0}\sum^{l}_{m=-l}P_{lm}Q\{j\}_{lm},
\end{equation}
where the functions $Q\{j\}_{lm}$ are given by (\ref{eq_3.82}).

The obvious properties of spherical functions, i.e,
\begin{equation}\label{eq_3.108}
Y^{l}{}_{m}(-\overline{n})=(-1)^{l}Y_{lm}(\overline{n}),\qquad 
\overline{Y^{l}{}_{m}}=Y^{l}{}_{-m}
\end{equation}
imply that $F$ is Hermitian in the sense of group algebra over ${\rm SU}(2)$ if and only if
\begin{equation}\label{eq_3.109}
\overline{P_{lm}}=(-1)^{l}\overline{P_{l(-m)}}.
\end{equation}

The Hermitian elements of the group algebra of ${\rm SU}(2)$ given by (\ref{eq_3.104}) are assumed to represent some important physical quantities. They have very suggestive tensorial structure and for $l=1$ they represent the angular momentum. Because of this there is a natural temptation to interpret them physically in terms of magnetic multipole momenta \cite{9}. Although in tensorial representation their system is redundant, it is convenient to expand with respect to them the density operators. The corresponding coefficients $P(l)^{a_{1}\ldots a_{l}}$ are directly related to the expectation values of multipoles, and it is reasonable to interpret them physically as magnetic polarizations of the corresponding order \cite{9}. It is clear that the physical situations, characterized by the fixed label $j$ of the Casimir invariant, possess multipole momenta and polarizations of the orders $l=0,1,\ldots,2j$. The algebraically non-redundant description of these objects is based on (\ref{eq_3.107})--(\ref{eq_3.109}).

Let us now discuss the quasiclassical limit. By this we mean the limit of "large quantum numbers" in equations like (\ref{eq_3.82}), (\ref{eq_3.83}) and others. An important aspect of this asymptotics is that the corresponding basic solutions are superposed with coefficients which are "slowly varying" functions of their arguments in some "wide" range of their values and practically vanishing outside this range. It is important that the range is simultaneously "wide" in the sense "much wider than one", but at the same time concentrated about some "large" mean value. This enables one to perform approximate "continuization" of discrete labels/(quantum numbers) and to replace their summation by integration.

For $l=0$ the substitution of 
\begin{equation}\label{eq_3.110}
f_{j0}=A\frac{\chi_{j0}}{\sin k/2}, \qquad A={\rm const},
\end{equation}
to (\ref{eq_3.83}) leads immediately to the following result:
\begin{equation}\label{eq_3.111}
f_{j0}=A\frac{\sin(2j+1)k/2}{\sin k/2}.
\end{equation}
But $\varepsilon(j)(0)=(2j+1)^{2}$, thus, $A=2j+1$, and finally
\begin{equation}\label{eq_3.112}
\varepsilon(j)(k)=(2j+1)\frac{\sin (2j+1)k/2}{\sin k/2}, \qquad \delta(k)=\sum_{j=0}^{\infty}(2j+1)\frac{\sin (2j+1)k/2}{\sin k/2}.\
\end{equation}
One can easily show that
\begin{equation}\label{eq_3.113}
f_{j,l+1}=\left(\frac{d}{dk}-\frac{l}{2}{\rm ctg}\frac{k}{2} \right)f_{j,l},
\end{equation}
therefore, iterating this recurrence formula one obtains the explicit formula for the multipole basis (\ref{eq_3.82}): 
\begin{equation}\label{eq_3.114}
f_{jl}=\prod_{n=l-1}^{0}\left(\frac{d}{dk}-\frac{n}{2}{\rm ctg}\frac{k}{2}\right)\varepsilon(j).
\end{equation}

Let us discuss the asymptotic expansions of such expressions in a domain $[0, a]$ where $a<2\pi$. One can show that for continuous functions $f$ on $[0, a]$ the following holds:
\begin{eqnarray}\label{eq_3.115}
&&\lim_{j \rightarrow \infty} \int_{0}^{a}f(k)\left(\frac{\sin(2j+1)k/2}{\sin k/2}-\frac{\sin(2j+1)k/2}{ k/2}\right)dk\nonumber\\
&&=
\lim_{j \rightarrow \infty} \int_{0}^{a}f(k)\frac{k/2-\sin k/2}{(k/2)\sin k/2}\sin(2j+1)\frac{k}{2}dk=0.
\end{eqnarray}
Incidentally, this statement is true for more general "sufficiently regular" functions $f$; they need not be continuous. Equation (\ref{eq_3.115}) means that in the integral mean-value sense in $[0, a]$ the functions $\varepsilon(j)$ with "sufficiently large" $j$-s  may be asymptotically replaced by
\begin{equation}\label{eq_3.116}
(2j+1)\frac{\sin(2j+1)k/2}{k/2}.
\end{equation}
And for "sufficiently large" values of $j$ the functions (\ref{eq_3.116}) are essentially concentrated about $k=0$.

Therefore, for any $\varepsilon>0$ there exists $n_{0}\in \mathbb{N}$ such that for any $n>n_{0}$ the following holds:
\begin{equation}\label{eq_3.117}
\left|\int_{0}^{a}f(k)\frac{\sin n k/2}{k/2}dk- \int_{0}^{\infty}f(k)\frac{\sin n k/2}{k/2}dk\right|<\varepsilon
\end{equation}
and 
\begin{equation}\label{eq_3.118}
\lim_{n\rightarrow \infty }\int_{0}^{a}f(k)\sin\frac{ n k/2}{k/2}dk=\pi f(0),
\end{equation}
i.e.,
\begin{equation}\label{eq_3.119}
\mathbb{N} \ni n \mapsto \frac{1}{\pi} \sin\frac{ n k/2}{k/2}  
\end{equation}
is a "Dirac-delta sequence".

The functions $\varepsilon(j)=(2j+1)\chi(j)$ are concentrated about $k=0$ and have there the maxima $(2j+1)^{2}$. At $k=2\pi$ they have the extrema $\pm (2j+1)^{2}$ depending on whether $j$ is respectively integer $(+)$ or half-integer $(-)$. For $j \rightarrow \infty$, $\varepsilon(j)$ may be replaced by
\begin{equation}\label{eq_3.120}
{}_{0}\varepsilon(j)=(2j+1)\frac{2}{k}\sin ((2j+1)k/2)
\end{equation}
in any interval $[0, a]$, $a<2\pi$. But it is also seen that $\varepsilon(j)$ may be replaced by
\begin{equation}\label{eq_3.121}
{}_{2\pi}\varepsilon(j)=\pm(2j+1)\frac{2}{2\pi-k}\sin ((2j+1)k/2)
\end{equation}
in any interval $[a, 2\pi]$, $a>0$; the signs $+/-$ appear respectively for integer/half-integer values of $j$. Therefore, globally, in the total ${\rm SU}(2)$-range $k\in [0, 2\pi]$, we have the following asymptotics for $j \rightarrow \infty$:
\begin{equation}\label{eq_3.122}
\varepsilon(j)\approx (2j+1)\sin (2j+1)k/2\left(\frac{2}{k}+(-1)^{2j}\frac{2}{2\pi-k}\right).
\end{equation}

\begin{center}
\includegraphics[scale=0.4]{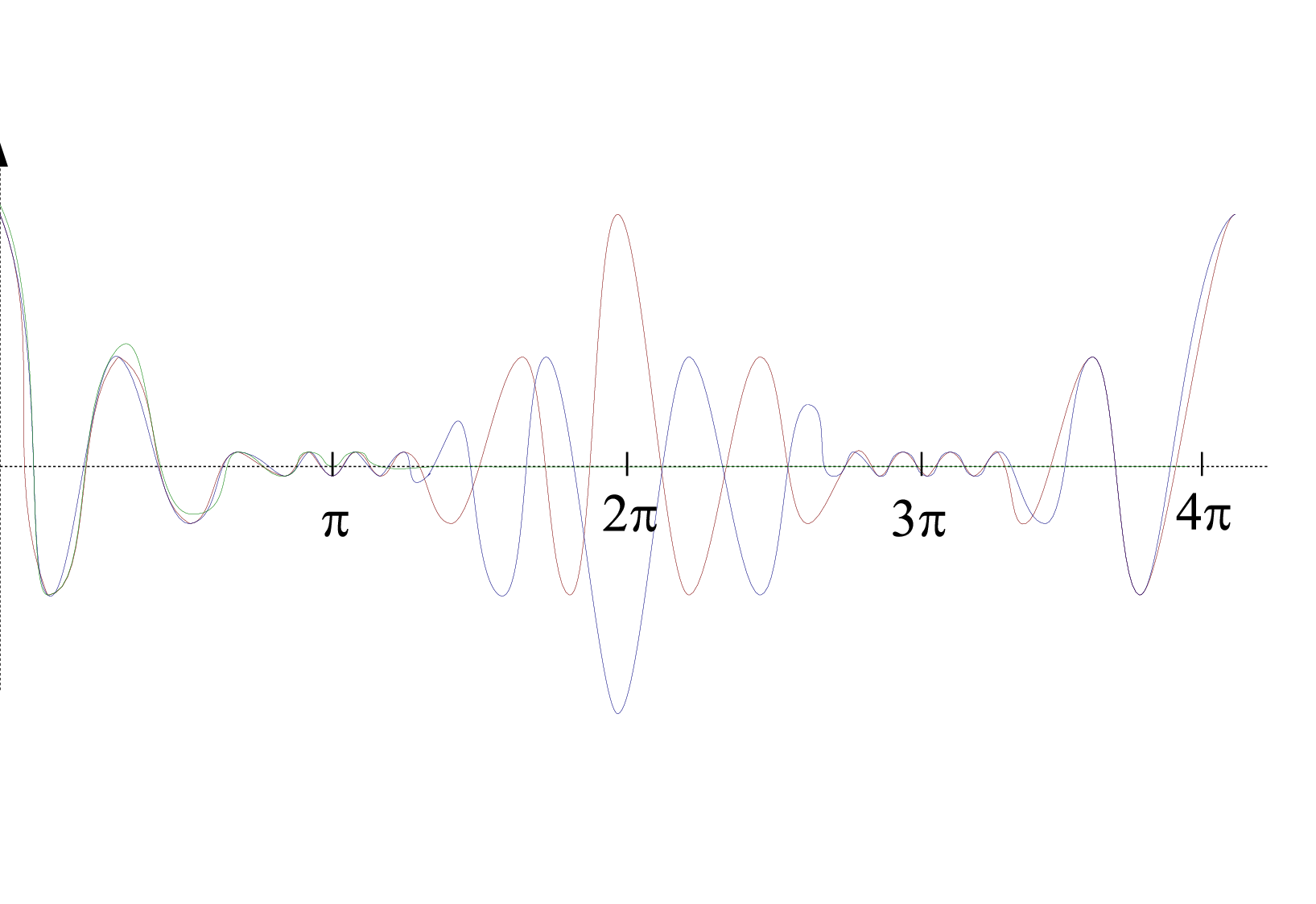}

{\rm Fig. 1}
\end{center}

The oscillating, sign-changing extremum at $k=2\pi$ is a purely quantum, spinorial effect. Such an effect does not appear on ${\rm SO}(3, \mathbb{R})$, when the range of $k$ is given by $[0, \pi]\subset \mathbb{R}$. However, when the functions $\varepsilon(j)$ are superposed with slowly-varying coefficients concentrated at large values of $j$, then the subsequent peaks approximately cancel each other. Nevertheless, for any fixed $j$, it does not matter how large one, we have the asymptotic formula (\ref{eq_3.122}) with both peaks; we shall write it symbolically:
\begin{equation}\label{eq_3.123}
\varepsilon(j)\approx \varepsilon_{0}(j)+\varepsilon_{2\pi}(j),
\end{equation}
where, obviously, $\varepsilon_{0}(j)$, $\varepsilon_{2\pi}(j)$ are concentrated respectively about $k=0$ and $k=2\pi$. The same is true for all other "radial" functions appearing in the multipole expansion (\ref{eq_3.82}).

Approximate equation for $f_{jl}$ about $k=0$ and for large values of $j$ has the following form:
\begin{equation}\label{eq_3.124}
\frac{d^{2}}{dk^{2}}{}_{0}f_{jl}+\frac{2}{k}\frac{d}{dk}{}_{0}f_{jl}+
\left(j(j+1)-\frac{l(l+1)}{k^{2}} \right){}_{0}f_{jl}=0.
\end{equation}
For ${}_{0}\varepsilon(j)={}_{0}f_{j0}$ one re-obtains the known expression:
\begin{equation}\label{eq_3.125}
{}_{0}\varepsilon(j)=(2j+1)\frac{\sin (2j+1)k/2}{k/2}.
\end{equation}
One can easily show that
\begin{eqnarray}
{}_{0}f_{j,l+1}&=&\left(\frac{d}{dk}-\frac{l}{k}\right){}_{0}f_{jl},
\label{eq_3.126}\\
{}_{0}f_{jl}&=&\left(\prod_{n=l-1}^{0}\left(\frac{d}{dk}-
\frac{n}{k}\right)\right){}_{0}\varepsilon(j)\label{eq_3.127}
\end{eqnarray}
in a complete analogy to (\ref{eq_3.113}), (\ref{eq_3.114}).

Another often used approximation for large $j$ is 
\begin{equation}\label{eq_3.128}
j(j+1)\mapsto \left(j+\frac{1}{2}\right)^{2};
\end{equation}
then the differential equation (\ref{eq_3.124}) becomes approximately
\begin{equation}
\frac{d^{2}}{dk^{2}}{}_{0}f_{jl}+\frac{2}{k}\frac{d}{dk}{}_{0}f_{jl}+
\left(\left(j+\frac{1}{2}\right)^{2}-\frac{l\left(l+1\right)}{k^{2}}\right)
{}_{0}f_{jl}=0.\label{eq_3.129}
\end{equation}

Again one can show that the approximate solutions of rigorous equations
for the large values of $j$ have the following form:
\begin{equation}
f_{jl}={}_{0}f_{jl}+{}_{2\pi}f_{jl},\label{eq_3.130}
\end{equation}
where
\begin{eqnarray}
{}_{0}f_{jl}&=&\left(\prod^{0}_{n=l-1}\left(\frac{d}{dk}-
\frac{n}{k}\right)\right){}_{0}\varepsilon\left(j\right),\label{eq_3.131}\\
{}_{2\pi}f_{jl}&=&\left(\prod^{0}_{n=l-1}\left(\frac{d}{dk}-
\frac{n}{2\pi-k}\right)\right){}_{2\pi}\varepsilon\left(j\right)\approx-\
{}_{2\pi}f_{j+\frac{1}{2},l}.\label{eq_3.132}
\end{eqnarray}

Obviously, $\varepsilon(j)$ in (\ref{eq_3.112}) has the profound geometric
interpretation of the generated unit of $M(j)$ and $\chi(j)=(1/(2j+1))\varepsilon(j)$
is the character of the $j$-th irreducible unitary representation
of ${\rm SU}(2)$. And seemingly, one might have an impression that the
asymptotic counterpart (\ref{eq_3.120}) is something "accidental",
non-interpretable in geometric terms. However, as a matter of fact,
it is an important object of the Fourier analysis on $\mathbb{R}^{3}\simeq {\rm SU}(2)'\simeq{\rm SO}(3,\mathbb{R})'$.

Indeed, it may be easily shown that the Fourier representation of
the Dirac delta distribution on $\mathbb{R}^{3}$ (as the Fourier
transform of unity), 
\begin{equation}
\delta\left(\overline{\omega}\right)=
\frac{1}{\left(2\pi\right)^{3}}\int\exp\left(i\underline{\varkappa}
\overline{\omega}\right)d_{3}\underline{\varkappa}
\label{eq_3.133}
\end{equation}
after performing the integration over angels becomes
\begin{equation}
\delta\left(\overline{\omega}\right)=
\frac{1}{\left(2\pi\right)^{3}}\int^{2\pi}_{0}
d\varphi\int^{\infty}_{0}d\varkappa\varkappa^{2}
\int^{\pi}_{0}d\vartheta\cos\vartheta\exp
\left(i\varkappa\omega\cos\vartheta\right),\label{eq_3.134}
\end{equation}
where, obviously, $\left(\varkappa,\vartheta,\varphi\right)$ are
spherical variables in the space $\mathbb{R}^{3}$ of vectors $\underline{\varkappa}$,
adapted to the direction of $\overline{\omega}$ as the "$z$-axis
direction". After the substitution $x=\cos\vartheta\in\left[1,-1\right]$
and $\varkappa=\zeta/2$ and elementary integrations, one
obtains the following formula: 
\begin{equation}
\delta\left(\overline{\omega}\right)=\frac{1}{16\pi^{2}}
\int^{\infty}_{0}d\zeta
\frac{\zeta\sin\left(\zeta\omega/2\right)}{\omega/2}
\label{eq_3.135}
\end{equation}
Under substituting $\zeta=(2j+1)$ it is turned into: 
\begin{equation}
\delta\left(\overline{\omega}\right)=\frac{1}{8\pi^{2}}
\int^{\infty}_{-1/2}dj\left(2j+1\right)
\frac{\sin\left(\left(2j+1\right)\omega/2\right)}{\omega/2}=\int dj\ \varepsilon_{\rm class}(j)(\omega).\label{eq_3.136}
\end{equation}
Expression
\begin{equation}
\varepsilon_{\rm class}(j)(\omega)=\left(2j+1\right)
\frac{\sin\left(\left(2j+1\right)\omega/2\right)}{\omega/2}
\label{eq_3.137}
\end{equation}
is an obvious counterpart of (\ref{eq_3.112}) and of its expression
for the Dirac distribution on ${\rm SU}(2)$, and the all other analogies
are easily readable. They are not merely formal analogies, the point
is that they are really true asymptotic approximations and geometric
counterparts. Discrete summation over the "quantum
number" $j$ is now replaced by the integration
over the continuous label $j$ corresponding to the non-compactness
of $\mathbb{R}^{3}\simeq{\rm SU}(2)'$ and well suited to the "classical"
nature of expressions. 

The superposed functions (\ref{eq_3.137})
play in the commutative group algebra of ${\rm SU}(2)'\approx\mathbb{R}^{3}$
the role of generating units of ideals $M(j)$ composed of functions
with the fixed "square of linear momentum"
$\left(j+1/2\right)^{2}\hbar^{2}\approx j\left(j+1\right)\hbar^{2}$;
the last approximate "equality"
corresponding to the "large"
values of $j$. This ideal is not minimal; the minimal ones just correspond
to the single exponents with the wave vectors $\underline{\varkappa}$,
i.e., "linear momenta" $\hbar\underline{\varkappa}$.
In $\varepsilon_{\rm class}(j)$ superposed are (with equal "coefficients")
all exponents $\exp\left(i\left(j+1/2\right)\overline{n}\cdot\overline{\omega}\right)$,
where $\overline{n}$ runs over the manifold $S^{2}(0,1)\subset\mathbb{R}^{3}$
of all unit vectors. The ideals $M(j)$ are minimal ones invariant
under the group ${\rm SO}(3,\mathbb{R})$ of outer automorphisms of ${\rm SU}(2)$.
Those outer automorphisms are not only algebraic automorphisms of
$\mathbb{R}^{3}$ as an Abelian additive group; in addition they preserve
the standard Euclidean metric in $\mathbb{R}^{3}$. This metric just
corresponds up to multiplicative constant factor to the Killing metric
of ${\rm SU}(2)'\approx\mathbb{R}^{3}$. Obviously, the above terms
like "group algebra" are now
used rather in a metaphoric sense, because we are dealing with continuous
spectrum and are outside of $L^{2}\left(\mathbb{R}^{3}\right)$ and $L^{1}\left(\mathbb{R}^{3}\right)$.
Everything may be rigorously formulated in terms of rigged Hilbert
spaces and direct integrals of Hilbert spaces, however, there is no
place for that here.

The above limit transition and asymptotics are meant in the sense
of truncation procedure in the rigorous group algebra of ${\rm SU}(2)$.

Quasiclassical limit is based on the truncation procedure of the group
algebra of ${\rm SU}(2)$. Namely, we take the subalgebra consisting of
all ideals $M(j)$ with $j\geq j_{0}$ for some fixed $j_{0}$, 
\begin{equation}
M\left(j\geq j_{0}\right):=\bigoplus_{j\geq j_{0}}M(j).
\label{eq_3.138}
\end{equation}
As mentioned, for large values of $j$, the generated units $\varepsilon(j)\in M(j)$
are essentially concentrated about $k=0$, $k=2\pi$, 
\begin{equation}
\varepsilon(j)\approx{}_{0}\varepsilon(j)+{}_{2\pi}\varepsilon(j),
\label{eq_3.139}
\end{equation}
cf. (\ref{eq_3.120}), (\ref{eq_3.121}), (\ref{eq_3.122}), (\ref{eq_3.123}),
and the following holds:
\begin{equation}
{}_{0}\varepsilon(j)(0)=\left(2j+1\right)^{2},\qquad
{}_{2\pi}\varepsilon(j)(2\pi)=\left(-1\right)^{2j}\left(2j+1\right)^{2}.
\label{eq_3.140}
\end{equation}
The larger truncation threshold $j_{0}$, the better the generated
unit of $M\left(j\geq j_{0}\right)$:
\begin{equation}
\varepsilon\left(j\geq j_{0}\right):=\sum^{\infty}_{j=j_{0}}\varepsilon(j)
\label{eq_3.141}
\end{equation}
is approximated by 
\begin{equation}
\varepsilon_{\rm class}\left(j\geq j_{0}\right):=\int^{\infty}_{j_{0}}dj\ \varepsilon_{\rm class}(j),
\label{eq_3.142}
\end{equation}
where $\varepsilon_{\rm class}(j)$ is given by (\ref{eq_3.137}). Of course, the
convergence of series (\ref{eq_3.141}) and integral (\ref{eq_3.142})
is meant in the distribution sense.

Projections of functions $A$, $B$ on ${\rm SU}(2)$ onto the truncated ideal
$M\left(j\geq j_{0}\right)$ will be denoted by 
\begin{equation}
\widetilde{A}=A\left(j \geq j_{0}\right), \qquad \widetilde{B}=B\left(j \geq j_{0}\right); \label{eq_3.143}
\end{equation}
the abbreviations $\widetilde{A}$, $\widetilde{B}$ are used when there is no danger of confusion.

For physically relevant functions $A$, $B$, the Peter-Weyl series expansions of $\widetilde{A}$, $\widetilde{B}$ may be reasonably approximated by continuous integral representations like (\ref{eq_3.136}), (\ref{eq_3.142}).

Let us go back to quantum states-densities represented in terms of non-redundant multipole expansions as follows:
\begin{eqnarray}
\varrho&=&\sum^{\infty}_{j=j_{0}} \sum^{2j}_{l=0} \sum^{l}_{m=-l} P(j)_{lm}Q\left\{j\right\}_{lm},\label{eq_3.144a}\\
Q\{j\}_{lm}\left(\overline{k}\right)&=& f_{jk}(k)Y_{lm}\left(\frac{\overline{k}}{k}\right),
\label{eq_3.144b}
\end{eqnarray}
where the expansion coefficients $P(j)_{lm}$ may be roughly interpreted as magnetic multipole moments. 

Quasiclassical states are represented by expressions (\ref{eq_3.144a}), (\ref{eq_3.144b}), where:
\begin{itemize}
\item $j_{0}$ is "large",

\item $P(j)_{lm}$ as functions of $j$ are concentrated in some ranges:
\begin{equation}
\left[\overline{j}- \Delta j/2, \overline{j}+ \Delta j/2 \right], \quad \overline{j} \gg \Delta j \gg 1, \label{eq_3.145}
\end{equation}

\item within this range, $P(j)_{lm}$ are slowly varying functions of $j$:
\begin{equation}
\left|P(j+1/2)_{lm}-P(j)_{lm}\right|\ll\left|P(j)_{lm}\right|. \label{eq_3.146}
\end{equation}
\end{itemize}

Algebraic operations of group algebra on ${\rm SU}(2)$ attain some very peculiar representation in quasiclassical limit in the above sense. So, let us write down the convolution formula for "truncated" functions:
\begin{eqnarray}
&&\left( A(j\geq j_{0}) \ast B(j\geq j_{0}) \right)\left(u\left(\overline{k}\right)\right) 
\nonumber\\
&&= \int A(j\geq j_{0})\left(u\left(\overline{l}\right)\right)B(j\geq j_{0})\left(u\left(-\overline{l}\right)u\left(\overline{k}\right)\right) \frac{4\sin^{2}\frac{l}{2}}{l^{2}} \frac{d_{3}\overline{l}}{16 \pi^{2}}. \label{eq_3.147}
\end{eqnarray}

The terms concentrated about $k=2\pi$, as it was seen, approximately cancel each other. One can assume that the integrated functions are essentially concentrated in a close neighbourhood of the unity in ${\rm SU}(2)$, i.e., the null of ${\rm SU}(2)'$. There, in the lowest order of approximation, we have
\begin{equation}
u\left(\overline{l}\right)u\left(\overline{k}\right) \approx u\left(\overline{l}+\overline{k}+ \frac{1}{2}\ \overline{l} \times \overline{k}\right). \label{eq_3.148}
\end{equation}

Performing the corresponding Taylor expansions in our integral formulas and making use of the earlier mentioned relationship between the variables $\overline{k}$ and $\overline{\omega}$, we finally obtain 
\begin{equation}
A(j \geq j_{0}) \ast_{{\rm SU}(2)} B(j \geq j_{0}) \approx A(j\geq j_{0}) \ast_{\mathbb{R}^{3}} B(j\geq j_{0}), \label{eq_3.149}
\end{equation}
where the convolution symbols on the left- and right-hand sides are respectively meant in the non-commutative ${\rm SU}(2)$-sense and commutative sense of $\mathbb{R}^{3} \simeq{\rm SU}(2)' \simeq{\rm SO}(3, \mathbb{R})'$. Obviously, (\ref{eq_3.149}) is meant in the sense of lowest-order approximation; the terms with higher-order derivatives are neglected. 

Similarly, for the quantum Poisson bracket we obtain the familiar expression:
\begin{equation}
\left[\widetilde{A},\widetilde{B}\right]= \frac{1}{i\hbar}\left(\widetilde{A} \ast_{{\rm SU}(2)}\widetilde{B} - \widetilde{B} \ast_{{\rm SU}(2)}\widetilde{A}\right) \approx \frac{1}{i\hbar}  \left(\left({\mathcal{A}}_{a}\widetilde{A}\right) \ast_{\mathbb{R}^{3}} \left(\omega^{a} \widetilde{B}\right)\right).\label{eq_3.150}
\end{equation}
Again we mean the lowest-order approximation, when the higher-derivatives terms following from the Taylor expansions are neglected. Obviously, as usual, $\mathcal{A}_{a}$ is the generator of inner automorphisms in ${\rm SU}(2)$, i.e., equivalently, of Killing rotations in ${\rm SU}(2)' \approx \mathbb{R}^{3}$,
\begin{equation}
\mathcal{A}_{a} = \varepsilon_{ab}{}^{c} \omega^{b} \frac{\partial}{\partial \omega^{c}}. \label{eq_3.151}
\end{equation}

So, in terms of Fourier representants $\widehat{\widetilde{A}}(\underline{\sigma})$, we have
\begin{eqnarray}
\left\{ \sigma_{i}, \sigma_{j} \right\} &=& \sigma_{k}\varepsilon^{k}{}_{ij},\label{eq_3.152a} \\
\left\{ \widehat{\widetilde{A}}, \widehat{\widetilde{B}} \right\} &=& \sigma_{k}\varepsilon^{k}{}_{ij} 
\frac{\partial \widehat{\widetilde{A}}}{\partial \sigma_{i}} \frac{\partial \widehat{\widetilde{B}}}{\partial \sigma_{j}}. \label{eq_3.152b}
\end{eqnarray}
In particular, for the evolution of density $\widetilde{\varrho}$ we obtain
\begin{equation}
\frac{\partial \widetilde{\varrho}}{\partial t} =\left[\widetilde{H},\widetilde{\varrho}\right]_{\mathbb{R}^{3}}, \qquad \frac{\partial \widehat{\widetilde{\varrho}}}{\partial t} = \left[\widehat{\widetilde{H}},\widehat{\widetilde{\varrho}}\right], \label{eq_3.154}
\end{equation}
where $H$ denotes the Hamiltonian. Taking appropriate Hamiltonians one obtains classical asymptotics of various dynamical models of the evolution of quantum angular momentum, or rather systems of quantum angular momenta. This includes complicated interactions between magnetic multipoles as described above.

Let finish with some comments concerning quasiclassical formulas which may be helpful when operating with some geometrically and physically important quantities.

First of all, let us observe that (\ref{eq_3.149}) is a merely zeroth-order approximation. The first-order approximation is given by
\begin{equation}
\widetilde{A} \ast_{{\rm SU}(2)} \widetilde{B} \approx \widetilde{A} \ast_{\mathbb{R}^{3}} \widetilde{B}+ \frac{1}{2}\left[\widetilde{A}, \widetilde{B}\right]_{\mathbb{R}^{3}}, \label{eq_3.155}
\end{equation}
where, let us remind $\left[\widetilde{A}, \widetilde{B}\right]_{\mathbb{R}^{3}}$ is the extreme right-hand side of (\ref{eq_3.150}). The second term is the lowest-order approximation to the ${\rm SU}(2)$-convolution commutator. It is well known that the commutator, or more precisely quantum Poisson bracket, describes infinitesimal transformations, in particular symmetries of quantum states (as described by density operators). It is well known that the operator eigenequation for density operators, 
\begin{equation}
\mathbf{A} \mathbf{\varrho}= a \mathbf{\varrho}, \label{eq_3.156}
\end{equation}
implies that the operators $\mathbf{A}$, $\mathbf{\varrho}$ do commute, thus, their quantum Poisson bracket vanishes,
\begin{equation}
\left\{\mathbf{A}, \mathbf{\varrho} \right\}_{Q}= \frac{1}{i\hbar}(\mathbf{A}\mathbf{\varrho}- \mathbf{\varrho}\mathbf{A})=0; \label{eq_3.157}
\end{equation}
obviously, it is assumed here that $\mathbf{A}$ represents a physical quantity, thus, it is self-adjoint, $\mathbf{A}^{+}=\mathbf{A}$. Therefore, the concept of eigenstate, in particular that of pure state (one satisfying a maximal possible system of compatible eigenconditions), unifies in a very peculiar way two logically distinct concepts: information and symmetry. Information aspect is that the physical quantity $\mathbf{A}$ has a sharply defined value on the state $\mathbf{\varrho}$; there is no statistical spread of measurement results. Symmetry aspect is that $\mathbf{\varrho}$ is invariant under the one-parameter group of unitaries, i.e., quantum automorphisms, generated by $\mathbf{A}$. On  the quantum level, symmetry properties are implied by information properties, because the quantum Poisson bracket is algebraically built of the associative product. This is no longer the case in quasiclassical limit and on the classical level, where the Poisson bracket and (commutative) associative product are algebraically independent on each other. But information and symmetry are qualitatively different things, therefore, on the quasiclassical level, the two first terms of the expansion for the non-commutative associative product should be taken into account when discussing classical concepts corresponding to eigenequations. Otherwise the physical interpretation of eigenconditions would be damaged.

Let us remind, following (\ref{eq_3.78a})--(\ref{eq_3.78c}), that differential equations satisfied by the functions $\varepsilon(j)_{mk}$ may be written in the following form:
\begin{eqnarray}
\mathbf{S}^{2}\ \varepsilon(j)_{mk} &=& j(j+1)\hbar^{2}\ \varepsilon(j)_{mk}, \label{eq_3.158a}\\
\mathbf{S}_{3}\ \varepsilon(j)_{mk} &=& m \hbar \ \varepsilon(j)_{mk}, \label{eq_3.158b}\\
\widehat{\mathbf{S}}_{3}\ \varepsilon(j)_{mk} &=& k \hbar \ \varepsilon(j)_{mk},\label{eq_3.158c}
\end{eqnarray}
with the known spectra of quantum numbers $j$, $m$, $k$. Rewriting these equations in terms of ${\rm SU}(2)$-convolutions we obtain
\begin{eqnarray}
\Sigma^{2} \ast \varepsilon(j)_{mk} &=& \varepsilon(j)_{mk} \ast \Sigma^{2} = j(j+1)\hbar^{2}\ \varepsilon(j)_{mk},\label{eq_3.159a}\\
\Sigma_{3} \ast \varepsilon(j)_{mk} &=& m \hbar\ \varepsilon(j)_{mk}, \label{eq_3.159b}\\
\varepsilon(j)_{mk} \ast \Sigma_{3} &=& k \hbar\ \varepsilon(j)_{mk},\label{eq_3.159c}
\end{eqnarray}
where, obviously, $\Sigma_{a}$ are given by (\ref{eq_3.97}) and $\Sigma^{2}$ denotes the convolution-squared vector $\Sigma_{a}$,
\begin{equation}
\Sigma^{2} = \Sigma_{1} \ast \Sigma_{1} + \Sigma_{2} \ast \Sigma_{2} +\Sigma_{3} \ast \Sigma_{3}. \label{eq_3.160}
\end{equation}
 
To obtain the quasiclassical counterparts of (\ref{eq_3.159a})--(\ref{eq_3.159c}) we must use the asymptotic formulas (\ref{eq_3.155}). It is more convenient to express them in terms of Fourier transforms, which were defined as functions on the Lie co-algebra $\left({\rm SU}(2)'\right)^{*} \simeq \mathbb{R}^{3}$. So, we shall use the coordinates $\sigma_{i}$ introduced above and the functions $\widehat{\varepsilon}(j)(\underline{\sigma})$ such that
\begin{equation}
\varepsilon(j)_{mn}(\overline{\varkappa})= \frac{1}{(2 \pi \hbar)^{3}} \int \widehat{\varepsilon}(j)_{mn}(\underline{\sigma}) \exp \left(\frac{i}{\hbar}\underline{\sigma} \overline{\varkappa} \right)d_{3}\underline{\sigma}. \label{eq_3.161}
\end{equation}
The left-hand sides of (\ref{eq_3.161}) are functions on the Lie algebra ${\rm SU}(2)' \simeq \mathbb{R}^{3}$ used to represent the approximate expressions for the elements of canonical basis (matrix elements of irreducible UNIREPS) as functions on ${\rm SU}(2)$. The system (\ref{eq_3.159a})--(\ref{eq_3.159c}) is expressed in terms of these Fourier transforms as follows:
\begin{eqnarray}
\sigma^{2}\ \widehat{\varepsilon}(j)_{mn}(\underline{\sigma}) &=& j(j+1)\hbar^{2}\ \widehat{\varepsilon}(j)_{mn}(\underline{\sigma}),\label{eq_3.162a}\\
\sigma_{3}\ \widehat{\varepsilon}(j)_{mn}(\underline{\sigma}) + \frac{1}{2}\left\{\sigma_{3}, \widehat{\varepsilon}(j)_{mn}(\underline{\sigma}) \right\} &=& m \hbar \ \widehat{\varepsilon}(j)_{mn}(\underline{\sigma}),\label{eq_3.162b}\\
\sigma_{3}\ \widehat{\varepsilon}(j)_{mn}(\underline{\sigma}) - \frac{1}{2}\left\{\sigma_{3}, \widehat{\varepsilon}(j)_{mn}(\underline{\sigma}) \right\} &=& n \hbar\ \widehat{\varepsilon}(j)_{mn}(\underline{\sigma}). 
\label{eq_3.162c}
\end{eqnarray}
The last two equations imply that
\begin{equation}
\left\{\sigma_{3}, \widehat{\varepsilon}(j)_{mn}(\underline{\sigma}) \right\} = (m-n) \hbar\ \widehat{\varepsilon}(j)_{mn}(\underline{\sigma}). \label{eq_3.163}
\end{equation}

It is convenient to use the polar angle $\varphi$ in the plane $\sigma_{3}=0$ of variables $\sigma_{1}$, $\sigma_{2}$ in ${\rm SU}(2)' \simeq \mathbb{R}^{3}$, 
\begin{equation}
{\rm tg}\ \varphi = \sigma_{2}/\sigma_{1}. \label{eq_3.164}
\end{equation}
Instead of cylindrical variables $\sigma_{3}$, $\varrho=\sqrt{\sigma_{1}{}^{2}+\sigma_{2}{}^{2}}$, $\varphi$ in the Lie co-algebra ${\rm SU}(2)'* \simeq \mathbb{R}^{3}$, we shall use the modified, spherically-cylindrical coordinates: 
\begin{equation}
\sigma=\sqrt{\sigma_{1}{}^{2}+\sigma_{2}{}^{2}+ \sigma_{3}{}^{2}}, \quad \sigma_{3}, \quad \varphi = {\rm arctg}\ \sigma_{2}/\sigma_{1}. \label{eq_3.165}
\end{equation}
They coincide with the canonical Darboux coordinates in ${\rm SU}(2)^{\prime\ast}$ as a Poisson manifold; their Poisson brackets have the following form:
\begin{equation}
\left\{\varphi, \sigma_{3} \right\}=1, \qquad \left\{ \sigma, \varphi \right\}=0, \qquad \left\{\sigma, \sigma_{3} \right\}=0. \label{eq_3.166}
\end{equation}
In particular, $\sigma^{2}= \underline{\sigma} \cdot \underline{\sigma}$ is the basic Casimir invariant. Its value surfaces $\sigma={\rm const}$ are canonically two-dimensional symplectic manifolds; the exceptional "s-state" value $\sigma=0$ is the singular co-adjoint orbit of dimension zero, just the origin of coordinates.

In these coordinates the following holds:
\begin{equation}
\left\{\sigma_{3}, f(\underline{\sigma}) \right\} = \frac{\partial}{\partial \varphi}f(\sigma, \sigma_{3}, \varphi), \label{eq_3.167}
\end{equation}
therefore, the system (\ref{eq_3.162a})--(\ref{eq_3.162c}) is solved as follows:
\begin{eqnarray}
&&\widehat{\varepsilon}(j)_{mn} = N(j) \delta \left(\sigma^{2}- \hbar^{2}j(j+1)\right)\delta\left(\sigma_{3}-\hbar \frac{m+n}{2} \right) \exp\left(i(m-n)\varphi\right)\qquad\ \label{eq_3.168} \\
&&= \frac{N(j)}{2\hbar \sqrt{j(j+1)}}\delta \left(\sigma- \hbar \sqrt{j(j+1)}\right)\delta \left(\sigma_{3}-\hbar \frac{m+n}{2} \right) \exp\left(i(m-n)\varphi\right),\nonumber
\end{eqnarray}
where $N(j)$ is a $j$-dependent normalization factor. It is defined by the 
demand that
\begin{equation}
\left.\varepsilon(j)_{mn}\right|_{\overline{\varkappa}=0}=(2j+1)\delta_{mn}. \label{eq_3.169}
\end{equation}

As already mentioned above, in quasiclassical situations the quantum-mys\-terious $j(j+1)$ is not very essential and may be replaced by $\left(j+ 1/2\right)^{2}$, or just by $j^{2}$.

Let us observe an interesting analogy with some formulas from the Weyl-Wigner-Moyal formalism for quantum systems with classical analogy. The "basis" of the wave function space consisting of non-normalizable, or rather "Dirac-$\delta$-normalized", states $\left|\underline{\pi}\right\rangle$ of fixed linear momentum $\underline{\pi}$ implies the following $H^{+}$-algebra "basis" in the space of phase-space functions (including the Moyal quasi-probability distributions):
\begin{equation}
\varrho_{\underline{\pi}_{1}, \underline{\pi}_{2}}\left(\overline{q},\underline{p}\right)= \delta\left(p-\frac{1}{2}\left(\underline{\pi}_{1}+ \underline{\pi}_{2}\right)\right) \exp\left(\frac{i}{\hbar}\left(\underline{\pi}_{1}- \underline{\pi}_{2}\right)\overline{q}\right). \label{eq_3.170}
\end{equation}
There is an obvious analogy with the term
\begin{equation}
\delta\left(\sigma_{3}-\frac{1}{2}\left(\mu_{1}+ \mu_{2}\right)\right) \exp\left(\frac{i}{\hbar}\left(\mu_{1}- \mu_{2}\right)\varphi\right) \label{eq_3.171}
\end{equation}
in (\ref{eq_3.168}), if we put $\mu_{1}=\hbar m$, $\mu_{2}=\hbar n$. This analogy is not accidental; however, there is no place here for more details.

Equation (\ref{eq_3.168}) and normalization condition (\ref{eq_3.169}) imply finally that
\begin{eqnarray}
\widehat{\varepsilon}(j)_{mn} &\approx& 16 \pi^{2}\hbar^{4}\left(j+\frac{1}{2}\right)^{2} \delta \left(\sigma^{2}- \hbar^{2}\left(j+\frac{1}{2}\right)^{2}\right)\nonumber\\
&&\delta\left(\sigma_{3}-\hbar \frac{m+n}{2} \right) \exp\left(i(m-n)\varphi\right),
\end{eqnarray}
therefore,
\begin{eqnarray}
\widehat{\mathcal{D}}(j)_{mn} &\approx& 8 \pi^{2}\hbar^{4}\left(j+\frac{1}{2}\right) \delta \left(\sigma^{2}- \hbar^{2}\left(j+\frac{1}{2}\right)^{2}\right)\nonumber\\
&&\delta\left(\sigma_{3}-\hbar \frac{m+n}{2} \right) \exp\left(i(m-n)\varphi\right),
\end{eqnarray}
with the same as previously, asymptotic "indifference" concerning $j(j+1)$, $\left(j+1/2\right)^{2}$, $j^{2}$ for large values of $j$.\bigskip

\noindent {\bf Warning:} It must be stressed that the above functions are not literally meant asymptotic expressions for $\varepsilon(j)_{mn}$, $\mathcal{D}(j)_{mn}$ for "large" values of $j$. They may be used instead of rigorous $\varepsilon(j)_{mn}$, $\mathcal{D}(j)_{mn}$ when superposing them with coefficients "slowly varying" as functions of $j$. And the very important point: The discrete quantum number $j$ may be then formally admitted to be a continuous variable and the summation with "slowly-varying" coefficients may be approximated by integration. In this way the compactness of ${\rm SU}(2)$ is seemingly "lost". This procedure is well known in practical applications of Fourier analysis, where often Fourier series may be approximated by Fourier transforms.

\section*{Acknowledgements}

This paper partially contains results obtained within the framework of the research project 501 018 32/1992 financed from the Scientific Research Support Fund in 2007-2010. The authors are greatly indebted to the Ministry of Science and Higher Education for this financial support.


\begin{thebibliography}{99}
\bibitem{12}
W. Ambrose, {\it Amer. Math. Soc.}, {\bf 57}, 364, 1945.

\bibitem{5}
L. E. Ballentine, {\it Quantum Mechanics: A Modern Development}, World Scientific Publishing Co. Ltd., Singapore-New Jersey-London-Hong Kong, 1998.

\bibitem{21}
A. A. Kirillov, {\it \'{E}l\'{e}ments de la Th\'{e}orie des Repr\'{e}sentations}, \'{E}ditions MIR, Moscow, 1974.

\bibitem{22}
A. A. Kirillov, {\it Merits and Demerits of the Orbit Method}, Bull. Amer. Math. Soc., {\bf 36}, 433--488, 1999.

\bibitem{20}
A. A. Kirillov, {\it Lectures on the Orbit Method}, Graduate Studies in Mathematics, {\bf 64}, American Mathematical Society, Providence, RI, 2004.

\bibitem{3}
L. D. Landau, E. M. Lifshitz, {\it Course of Theoretical Physics. Vol. III. Quantum Mechanics}, Pergamon Press, London, 1958.

\bibitem{2a}
L. H. Loomis, {\it An Introduction to Abstract Harmonic Analysis}, D. Van Nostrand Company, Inc., Princeton-New Jersey-Toronto-London-New York, 1953.

\bibitem{1a}
G. W. Mackey, {\it The Mathematical Foundations of Quantum Mechanics}, Benjamin, New York, 1963.

\bibitem{11}
K. Maurin, {\it General Eigenfunction Expansions and Unitary Representations of Topological Groups}, PWN --- Polish Scientific Publishers, Warsaw, 1968.

\bibitem{13}
K. Maurin, {\it Methods of Hilbert Spaces}, PWN --- Polish Scientific Publishers, Warsaw, 1972.

\bibitem{14}
K. Maurin, {\it Analysis. Part I--III}, D. Reidel-PWN, Dordrecht-Warszawa, 1980.

\bibitem{10}
L. Pontryagin, {\it Topological Groups}, Princeton University Press, Princeton, New Jersey, 1956.

\bibitem{4}
M. E. Rose, {\it Elementary Theory of Angular Momentum}, Dover Publications, 1965.

\bibitem{Rudin}
W. Rudin, {\it Fourier Analysis on Groups}, Interscience Publ., New York-London, 1962.

\bibitem{8}
F. E. Schroeck, Jr., {\it Quantum Mechanics on Phase Space}, Kluwer Academic Publishers, Dordrecht-Boston-London, 1996.

\bibitem{17}
J. J. S\l awianowski, {\it Abelian Groups and the Weyl Approach to Kinematics. Non-Local Function Algebras}, Reports on Mathematical Physics, {\bf 5}, no. 3, 295--319, 1974.

\bibitem{18}
J. J. S\l awianowski, {\it Geometry of Phase Spaces}, PWN --- Polish Scientific Publishers, Warsaw; John Wiley \& Sons, Chichester-New York-Brisbane-Toronto-Singapore, 1991.

\bibitem{16}
J. J. S\l awianowski, V. Kovalchuk, {\it Schr\"{o}dinger and Related Equations as Hamiltonian Systems, Manifolds of Second-Order Tensors and New Ideas of Nonlinearity in Quantum Mechanics}, Reports on Mathematical Physics, {\bf 65}, no. 1, 29--76, 2010; arXiv:0812.5055.

\bibitem{19}
J. J. S\l awianowski, V. Kovalchuk, B. Go\l ubowska, A. Martens, E. E. Ro\.{z}ko, {\it Quantized Excitations of Internal Affine Modes and Their Influence on Raman Spectra}, Acta Physica Polonica B, {\bf 41}, no. 1, 165--218, 2010; arXiv:0901.0243.

\bibitem{1} J. J. S\l awianowski, V. Kovalchuk, A. Martens, B. Go\l ubowska, E. E. Ro\.{z}ko, {\it Quasiclassical and Quantum Systems of Angular Momentum. Part I. Group Algebras as a Framework for Quantum-Mechanical Models with Symmetries}, arXiv:1007.4121.

\bibitem{2} J. J. S\l awianowski, V. Kovalchuk, A. Martens, B. Go\l ubowska, E. E. Ro\.{z}ko, {\it Quasiclassical and Quantum Systems of Angular Momentum. Part II. Quantum Mechanics on Lie Groups and Methods of Group Algebras}, arXiv:1008.0512.

\bibitem{9}
W. M. Tulczyjew, {\it The Theory of Systems with Internal Degrees of Freedom}, Lecture Notes, Department of Physics, Lehigh University, Bethlehem, Pennsylvania, 1964.

\bibitem{6}
H. Weyl, {\it The Theory of Groups and Quantum Mechanics}, Dover, New York, 1950.

\bibitem{15}
H. Weyl, {\it Symmetry}, Princeton University Press, Princeton, New Jersey, 1952.

\bibitem{7}
E. P. Wigner, {\it Gruppentheorie und Ihre Anwendungen auf die Quantenmechanik der Atomspektren}, Vieweg Verlag, Braunschweig, 1931. English Translation by J. J. Griffin, {\it Group Theory and its Application to the Quantum Mechanics of Atomic Spectra}, Academic Press, New York, 1959.

\end{thebibliography}
\end{document}